\documentclass[11pt,a4paper]{article}
\usepackage[T1]{fontenc}
\usepackage{lmodern}
\usepackage[margin=27mm,headheight=14pt]{geometry}
\usepackage{amsmath,amssymb,amsthm,mathtools}
\usepackage{microtype,graphicx,booktabs,longtable,array,pdflscape}
\usepackage[dvipsnames]{xcolor}
\usepackage{enumitem,fancyhdr}
\usepackage{needspace}
\usepackage{listings}
\usepackage[numbers,sort&compress]{natbib}
\usepackage[colorlinks=true,linkcolor=MidnightBlue,citecolor=MidnightBlue,urlcolor=MidnightBlue]{hyperref}
\usepackage{xurl}
\usepackage[capitalise,nameinlink,noabbrev]{cleveref}
\usepackage[font=small,labelfont=bf]{caption}
\hypersetup{pdftitle={Trotter Error Bounds in Lean 4 with a Cancellation-Aware Analysis of the Fourth-Order Suzuki Formula},pdfauthor={Bingyu Xia, Jue Xu, and Yue Wang},pdfsubject={Product formulas and Hamiltonian simulation}}
\setlist{itemsep=3pt,topsep=5pt}
\numberwithin{equation}{section}
\newtheorem{theorem}{Theorem}[section]
\newtheorem{proposition}[theorem]{Proposition}
\newtheorem{lemma}[theorem]{Lemma}
\newtheorem{corollary}[theorem]{Corollary}
\theoremstyle{definition}\newtheorem{definition}[theorem]{Definition}
\theoremstyle{remark}
\newcommand{\norm}[1]{\left\lVert#1\right\rVert}
\newcommand{\abs}[1]{\left|#1\right|}
\newcommand{\ad}{\operatorname{ad}}
\newcommand{\sinc}{\operatorname{sinc}}
\newcommand{\Acal}{\mathcal A}
\newcommand{\Rcal}{\mathcal R}

\newcommand{\fqfplink}[2]{\href{https://github.com/BryceT233/Foresight-Quantum-Formalization-Projects#1}{#2}}
\newcommand{\bp}{B_J}
\newcommand{\bs}{B_{\mathrm{single}}}
\newcommand{\bc}{B_{\mathrm{cancel}}}
\newcommand{\bo}{B_{\mathrm{old}}}
\newcommand{\field}{\mathbb Q[u]/(60u^3-48u^2+12u-1)}

\title{Trotter Error Bounds in Lean~4\\
with a Cancellation-Aware Analysis of the\\
Fourth-Order Suzuki Formula}
\author{Bingyu Xia\textsuperscript{2}, Jue Xu\textsuperscript{1}, and Yue Wang\textsuperscript{1,2}\thanks{Corresponding author.}\\[0.5em]
  \small\textsuperscript{1}QICI Quantum Information and Computation Initiative,\\
  \small Department of Computer Science, School of Computing and Data Science,\\
  \small The University of Hong Kong, Pokfulam Road, Hong Kong SAR, China\\[0.2em]
  \small\textsuperscript{2}Foresight Quantum, Shanghai, China}
\date{19 November 2026}
\begin{document}
\maketitle
\begin{abstract}
We formalize the commutator-scaling bound of Childs et al.\ in Lean~4.
For skew-adjoint generators, the local order condition gives an explicit
integer step count sufficient to meet a prescribed error tolerance.
We apply the theorem in Lean to the binary fourth-order Suzuki formula.
For this formula we can also account for cancellation within and between
steps. Using local remainder estimates of Casas and Murua, we sum the
signed quintic and centered seventh-degree error terms after conjugating
each to the interaction picture. For every positive integer step count,
we bound the difference between this sum and the global error operator.
At fixed time the explicit remainder is eighth order in the step size,
and the integral of the conjugated quintic term gives the exact leading
error coefficient. The Lean proofs cover the two-generator finite-sum
estimates, including the spectral norm for complex matrices.

On five ten-site nearest-neighbor Heisenberg instances at time $10$ and
tolerance $10^{-3}$, the transported sum meets the tolerance at a mean
of $129.4$ steps. Taking a separate norm at each step with the same
polynomial and remainder gives a mean first crossing between $184.6$
and $185.2$ steps. The accumulated Casas--Murua local bound at $q=8$
gives a mean of $184.8$ steps.
\end{abstract}
\section{Introduction}\label{sec:scope}
How many product-formula steps suffice to simulate a Hamiltonian at
a given time and accuracy? Each step requires a fixed number of
exponentials, and two error bounds of the same order can give quite
different step counts.
On five ten-site Heisenberg fields we hold the local error polynomial
and its remainder fixed. Summing the conjugated errors before taking
their norm reduces the mean step count by about $30\%$ compared with
taking a norm at each step (\cref{tab:controlled-transport}).

Consider a Hamiltonian $H=\sum_j H_j$ with terms that can be
exponentiated individually. A staged product of exponentials approximates
$e^{-itH}$. Its error depends on which terms commute, how they are
ordered, and how the error from one step evolves during subsequent
steps. The commutator-scaling theory of \citet{childs} controls an
order-$p$ formula by a sum of norms of nested commutators of length
$p+1$. Vanishing commutators drop out of the sum, and the bound applies to an
entire class of formulas. For a particular formula, however, summing
commutator norms discards the relative signs of the coefficients.
Summing one-step error norms also discards the relative orientations
of errors from different steps.

Our Lean~4 formalization derives an explicit sufficient step count
from the general bound (\cref{thm:general-steps}). It specifies the
product order and the local order condition at zero, and proves the
counting inequality that replaces ordered conjugations by a sum over
Hamiltonian terms. We apply the theorem in Lean to Suzuki's five-stage
formula \cite{suzuki}.

For this product the signed leading error is a degree-five Lie
polynomial $L_5$. Centering the one-step error gives a seventh-degree
correction $N_7$. We transport both operators into the interaction
picture and take one norm after summing over all steps. At fixed $t$
the difference between the sum and the global error operator has an
explicit bound of order $(t/r)^8$ (\cref{thm:main-septic}). The quintic
term determines the exact coefficient of $r^{-4}$
(\cref{thm:main-global}).

\paragraph{Comparison with prior bounds.}
For the same binary formula, \citet[Proposition~J.1]{childs} bound the
error by eight weighted commutator norms. We compare those weights
with the signed quintic polynomial in \cref{thm:main-local}.
\citet[Eq.~(35)]{casas2026} analyze this five-block composition.
Their Proposition~4.7 and Remark~4.8 give local defect estimates at
different truncation orders, and their Section~2.4 accumulates the local
bound over the steps. We use this local analysis, the symmetric defect
of \citet{auzinger}, and Lie-series methods \cite{Somma2016,casas}.
The additional calculation identifies the seventh-degree coefficient
of the centered step and propagates it together with the quintic term.

Error interference and its dependence on eigenspaces have been studied
by \citet{tran,chen}. In particular, \citet[Eq.~(2) and Appendix~A]{chencommutant}
retains the linear sum of conjugated one-step errors and derives a
quadratic remainder estimate when $r$ times the one-step error norm is
less than one. Our contraction argument bounds the nonlinear remainder
by $\binom r2\delta^2$ for every positive $r$. Replacing the one-step
error with its centered polynomial costs at most another $r\eta$
(\cref{prop:centered-global-transfer}). \citet[Section~III.1]{maxwell2026}
study the energy-diagonal part of the leading BCH error in a joint
large-time, small-step limit. Our fixed-time coefficient retains the
off-diagonal blocks with their finite-time phase integrals
(\cref{thm:main-global}). \citet{liu2026} use local ordering information
to evaluate and accumulate single-step bounds on long chains. Our
numerical comparison evaluates the transported operator itself on small
chains where its spectral norm is accessible.

\paragraph{Where to find the Lean proofs.}
The proofs are available in \fqfplink{}{Foresight Quantum Formalization
Projects (FQFP)}~\cite{FQFP}, a Lean~4 library maintained by a research
group at Foresight Quantum. Bingyu Xia started the project to formalize
categorical quantum mechanics. The proofs reported here use Lean~4.36.0
and Mathlib~v4.36.0~\cite{MathlibCPP}. The product-formula development
incorporates results from Jue Xu's Lean-BCH
and Lean-Trotter repositories~\cite{LeanBCH,LeanTrotter}.
\fqfplink{/blob/e009b1d262780b7414e1d35afcf8b8670fd4beac/FQFP/Analysis/Normed/Algebra/ProductFormula/ErrorBounds.lean}{\texttt{ErrorBounds.lean}}
contains the general commutator bound, and
\fqfplink{/blob/e009b1d262780b7414e1d35afcf8b8670fd4beac/FQFP/Analysis/Normed/Algebra/ProductFormula/SuzukiSepticGlobalError.lean}{\texttt{SuzukiSepticGlobalError.lean}}
contains the global bound for the centered Suzuki error.
\Cref{app:childs,app:lean} map the results to Lean declarations and
explain how to reconstruct the source version used here.

\section{Verified commutator bounds and sufficient step counts}
\label{sec:general}

\subsection{Ordered products and local order}
For a finite-dimensional Hamiltonian $H=\sum_{j=1}^{m}H_j$ on a complex
Hilbert space, set $X_j=-iH_j$ and $G=\sum_j X_j$. We use the spectral
norm. Thus $e^{sX_j}$ and $e^{sG}$ are unitary for real $s$. We also
use the generator notation $X_j$ for arbitrary elements when discussing
a general normed algebra.

A staged formula $P$ has $\Upsilon$ stages. Stage $v$ chooses a
permutation $\pi_v$ of the $m$ summands and real numbers
$|a_{vj}|\le1$. Writing $A_{vj}(s)=\exp(s a_{vj} X_{\pi_v(j)})$, set
\[
 P_v(s)=A_{vm}(s)\cdots A_{v1}(s),\qquad
 P(s)=P_\Upsilon(s)\cdots P_1(s).
\]
The earliest factor acts on the right. The local order hypothesis means
$\norm{P(s)-e^{sG}}=O(|s|^{p+1})$ as $s\to0$ for the given generators.
For $[U,V]=UV-VU$ define
\begin{equation}
 \alpha_p(X)=\sum_{j_0,\ldots,j_p=1}^{m}
 \norm{\ad_{X_{j_p}}\cdots\ad_{X_{j_1}}X_{j_0}},
 \qquad \ad_U V=[U,V].
 \label{eq:alpha-general}
\end{equation}
The indices range independently, including repetitions. This sum depends
on the generators and $p$ alone. The coefficients and stage permutations
enter when we pass from the derivative of $P$ to this sum.

\begin{theorem}[Commutator bound and step count]\label{thm:general-steps}
Let the $X_j$ be skew-adjoint, $p\ge1$, and $\Upsilon\ge1$. If $P$ has
local order $p$, put $C_p=2\Upsilon^{p+1}\alpha_p(X)/(p+1)$.
For $t\ge0$ and integers $r\ge1$,
\begin{equation}
 \norm{P(t/r)^r-e^{tG}}\le C_p\,t^{p+1}/r^p.
 \label{eq:general-steps}
\end{equation}
For any tolerance $\varepsilon>0$ it suffices to choose
$r\ge\max\{1,\lceil(C_p t^{p+1}/\varepsilon)^{1/p}\rceil\}$.
\end{theorem}

\begin{proof}
Use the additive kernel $T(s)=P(s)^{-1}(P'(s)-GP(s))$. Variation of
constants gives
\begin{equation}
 P(h)-e^{hG}=\int_0^h e^{(h-s)G} P(s)T(s)\,ds .
 \label{eq:general-duhamel}
\end{equation}
We use the finite-conjugation estimate of \citet{childs}, keeping track
of its constants. List the summands in product order as $Z$, bounding
the stage coefficients by $|a_{vj}|\le1$. For an endpoint $W$ and a
list of length $b$, sum the norms of
$\ad_{Z_b}^{q_b}\cdots\ad_{Z_1}^{q_1} W$ over $|q|=p$ with weights
$p!/\prod_i q_i!$.
Call this ordered sum $A_p(Z,W)$, using the conjugation order of the
expansion. The finite expansion of $T$ and the local order hypothesis
cancel the lower Taylor coefficients. The conjugation remainder for
skew-adjoint generators gives
\[
 \norm{T(s)}\le
 \frac{2\Upsilon |s|^p}{p!}\sum_j A_p(Z,X_j),\qquad
 \sum_j A_p(Z,X_j)\le p!\,\Upsilon^p\,\alpha_p(X).
\]
For the second inequality, each of the $p$ bracket entries has at most $\Upsilon$
stage placements for a fixed generator index, and the multinomial
weight is bounded by $p!$. \Cref{app:childs} gives the Lean declarations for this counting argument
and the kernel estimate. Multiplication by the unitaries in
\cref{eq:general-duhamel} preserves the norm, so integration from $0$
to $h$ gives the one-step bound $C_p h^{p+1}$. Telescoping
$P(h)^r-(e^{hG})^r$ multiplies that bound by at most $r$.
Setting $h=t/r$ gives \cref{eq:general-steps}. Solving for the least
positive integer satisfying the resulting sufficient condition gives
the stated threshold.
\end{proof}

\subsection{Error representations and convergence rates}
The formalization checks three representations of the same error.
The additive kernel is $T$ above. The correction to the instantaneous
generator is $E(s)=P'(s)P(s)^{-1}-G$, so $P'=(G+E)P$.
The multiplicative error $M(s)=e^{-sG}P(s)-I$ is the interaction-picture
error for one step. The four local order conditions
\[
 P(s)-e^{sG}=O(|s|^{p+1}),\quad T(s)=O(|s|^p),\quad
 E(s)=O(|s|^p),\quad M(s)=O(|s|^{p+1})
\]
are equivalent for these products. Integration accounts for the change
in degree between the kernels and the errors. In the skew-adjoint case
$\norm{M(s)}=\norm{P(s)-e^{sG}}$. In a complete unital real normed
algebra with $\norm{1}=1$, for arbitrary generators of a formula of
order $p\ge1$ with $\Upsilon>0$, both single-step errors are bounded
for $h\ge0$ by
\[
 C_p h^{p+1}\exp\!\left(2h\Upsilon\sum_j\norm{X_j}\right).
\]
The commutator step-count theorem displays the dependence on time,
tolerance and the summands. The separate one-norm scaling declaration
gives $O(r^{-p})$ for fixed summands, order, formula and time.

Low-order formulas show how a specific ordering can improve the
constant. Write $U_j=\sum_{k>j}X_k$ and let $E_1(t,r)$ and $E_2(t,r)$
denote the global norm errors of the right-to-left Lie--Trotter
product and its symmetric Strang counterpart. For skew-adjoint
generators the verified prefactor bounds give
\begin{align*}
 E_1(t,r)&\le \frac{t^2}{2r}\sum_j\norm{[U_j,X_j]},\\
 E_2(t,r)&\le \frac{t^3}{r^2}\sum_j
 \left(\frac{\norm{[U_j,[U_j,X_j]]}}{12}
       +\frac{\norm{[X_j,[X_j,U_j]]}}{24}\right).
\end{align*}
The local order of these finite products gives fixed-time
rates $O(r^{-1})$ and $O(r^{-2})$ even for arbitrary bounded generators
in a complete real normed algebra with normalized unit. The binary
five-stage composition in \cref{eq:s4} has rate $O(r^{-4})$ in the
same setting. These rates hold for each fixed real $t$, including
negative times. The finite-step refinements below use unitarity.

\subsection{The binary Suzuki formula as a staged product}
\label{sec:stage-bridge}
Let $Z=(X,Y)$ be a pair of skew-adjoint generators and take
$P=\texttt{suzukiData}\ 2\ 2$. Each Strang block occupies two stages in
the recursive encoding, one for each ordering of the pair. Five blocks
therefore give $\Upsilon=10$. Combining adjacent exponentials of the same
generator gives the eleven factors of $S(h)$ in \cref{eq:s4}. The file
\path{formalization/SuzukiStageBridge.lean} verifies both identifications
needed to apply \cref{thm:general-steps}:
\begin{gather*}
 (\texttt{suzukiData}\ 2\ 2).\texttt{eval}\ Z\ h
   =\texttt{fiveStageStrangProduct}\ X\ Y\ \texttt{suzukiP}\ h,\\
 (\texttt{suzukiData}\ 2\ 2).\texttt{IsOrderOf}\ 4\ Z .
\end{gather*}
The first identity matches the two product conventions. The order proof
uses the quintic expansion of the same product at zero. Substituting
these facts into \path{trotter_error_bound_comm_scaling_pow} gives,
for $t\ge0$ and $r\ge1$,
\begin{equation}
 \norm{S(t/r)^r-e^{t(X+Y)}}\le
       40000\,\alpha_4(Z)\,\frac{t^5}{r^4},
 \qquad 40000=\frac{2\cdot10^5}{5}.
 \label{eq:binary-general}
\end{equation}
For the same product, the eight weights of
\citet[Proposition~J.1]{childs} give the formula-specific scalar bound
$B_Jt^5/r^4$. Both bounds sum commutator norms. To recover the signs
within a step and the orientation of successive errors, we return to
$S(h)-e^{h(X+Y)}$ before taking these norms. Its centered expansion
supplies the two operators propagated in \cref{thm:main-septic}.

\section{The fourth-order Suzuki bound}
\label{sec:results}
\subsection{The signed local error}
Let $H=A+B$, where $A,B$ are Hermitian matrices on a finite-dimensional
complex Hilbert space. Use the spectral norm and put $X=-iA$, $Y=-iB$
and $G=X+Y=-iH$. Write $[P,Q]=PQ-QP$ and $\ad_P(Q)=[P,Q]$.
One Suzuki step is
\begin{align}
 u&=(4-\sqrt[3]{4})^{-1},\qquad v=1-4u,\label{eq:u}\\
 S_2(h)&=e^{hX/2}e^{hY}e^{hX/2},\qquad
 S(h)=S_2(uh)^2S_2(vh)S_2(uh)^2.\label{eq:s4}
\end{align}
This step is unitary and symmetric: $S(-h)=S(h)^{-1}$.
For total time $t>0$ and an integer $r\ge1$, set $h=t/r$ and
$\epsilon_r(t)=\norm{S(h)^r-e^{tG}}$. All asymptotic statements below
hold with the generators and $t$ fixed. Write $L_5$ for the exact
degree-five local error polynomial, equivalently the coefficient of
$h^5$ in the formal logarithm of $S(h)$, and $\ell=\norm{L_5}$.
The five Strang blocks produce eleven exponentials after adjacent factors
are merged. The relation $4u^3+v^3=0$ cancels the cubic error, and the
palindrome eliminates even degrees of the formal logarithm.
The six Hall coefficients of $L_5$ are given in \cref{tab:hall}.

\begin{corollary}[Local specialization with the complete leading polynomial]
\label{thm:main-local}
For every real $h$,
\begin{equation}
 \norm{S(h)-e^{hG}}\le |h|^5\ell+|h|^7C_7,
 \qquad
 \lim_{h\to0^+}h^{-5}\norm{S(h)-e^{hG}}=\ell.
 \label{eq:main-local}
\end{equation}
Here $C_7$ is the explicit nonnegative commutator sum in \cref{eq:c7}.
The same inequality holds with the grouped constant
$C_7^{\mathrm g}$ in \cref{eq:refined-constants}.
The original constant uses the cited finite-conjugation estimate;
the grouped and integral forms are Casas--Murua specializations in \cref{sec:cm}.
\end{corollary}

For this composition, \citet[Proposition J.1]{childs} give the bound
$|h|^5\bp$, where $\bp=\sum_w b_w\norm{R_w}$ sums eight weighted
commutator norms. To obtain $\ell=\norm{\sum_w d_wR_w}$, we add the signed terms of
$L_5$ before taking their norm.
At the standard Suzuki parameter, the componentwise rational comparison
in \cref{eq:quintic-eighth} gives $\ell\le\bp/8$.
With the seventh-order remainder included, the original local bound is
at most the published estimate whenever
$|h|\max(\norm{X},\norm{Y})\le0.97$. \Cref{sec:local} gives the
rational-interval calculation for this condition. The grouped estimate
uses a separate constant $C_7^{\mathrm g}$, whose integral counterpart
satisfies $C_7^{\mathrm{CM}}\le C_7^{\mathrm g}$ (\cref{sec:cm}).

\subsection{A finite-step enclosure for the transported error}
For the global error define the continuous and discrete transported sums
\begin{align}
 K(t)&=\int_0^t e^{-sG}L_5e^{sG}\,ds,\label{eq:main-k}\\
 K^{\mathrm d}_{t,h}(B)&=h\sum_{k=0}^{r-1}
 e^{-(k+1/2)hG}B e^{(k+1/2)hG}.\label{eq:main-kd}
\end{align}
Define $N_7$ by the centered expansion
\begin{equation}
 e^{-hG/2}(S(h)-e^{hG})e^{-hG/2}
       =h^5L_5+h^7N_7+O(h^9).
\end{equation}
If $L_7$ is the degree-seven formal logarithm coefficient, then
$N_7=L_7+[G,[G,L_5]]/24$.
For the defect coefficient $M_7$ in \cref{eq:m7}, set
$\mu=\norm{M_7}$ and, for $h\ge0$,
\begin{align}
 \delta_9(h)&=h^5\ell+h^7\mu+h^9C_9,\label{eq:main-delta9}\\
 \eta_9(h)&=h^9D_9+\tfrac12h^{10}\ell^2+h^{12}\ell\mu
   +\tfrac{h^{14}}{14}(5\ell C_9+7\mu^2)
   +\tfrac7{16}h^{16}\mu C_9.\label{eq:main-eta9}
\end{align}
The finite grouped constants $C_9,D_9$ are specified in
\cref{eq:refined-constants,eq:d9}.

\begin{theorem}[Finite-step Suzuki error in the interaction picture]
\label{thm:main-septic}
For all $t>0$ and integers $r\ge1$, put
\begin{equation}
 \Lambda_{t,h}=h^4K^{\mathrm d}_{t,h}(L_5)
                   +h^6K^{\mathrm d}_{t,h}(N_7),\qquad
 R_7(t,h)=r\eta_9(h)+\binom r2\delta_9(h)^2.
\end{equation}
Then
\begin{align}
 \norm{e^{-tG}S(h)^r-I-\Lambda_{t,h}}&\le R_7(t,h),\label{eq:main-septic}\\
 \abs{\epsilon_r(t)-\norm{\Lambda_{t,h}}}&\le R_7(t,h),
 \qquad R_7(t,h)=O(h^8).
\end{align}
\end{theorem}
The Suzuki formula has order four. The $O(h^8)$ estimate bounds what
remains after subtracting the two transported terms. Taking one norm
of their sum retains cancellation both between degrees and between
steps. Both the original and refined estimates include all nonlinear
product terms and hold for every positive step size.
Proofs of \cref{thm:main-local,thm:main-global,thm:main-septic} are
given in \cref{sec:local,sec:global,sec:septic}, respectively.

\subsection{The sharp coefficient at fixed time}
Let $\delta(h)$ and $\eta(h)$ be the explicit local distance and
centered remainder bounds in \cref{eq:delta,eq:eta}, and set
\begin{equation}
 R_5(t,h)=r\eta(h)+\binom r2\delta(h)^2
       +\frac{th^6}{24}\norm{[G,[G,L_5]]}.
 \label{eq:main-r5}
\end{equation}
This is exactly $\Rcal(t,h)$ in \cref{eq:global-rem}.

\begin{theorem}[Global cancellation and sharp coefficient]
\label{thm:main-global}
For all $t>0$ and integers $r\ge1$,
\begin{equation}
 \norm{e^{-tG}S(h)^r-I-h^4K(t)}\le R_5(t,h),
 \qquad R_5(t,h)=O(h^6).
 \label{eq:main-global}
\end{equation}
In particular,
\begin{equation}
 \abs{\epsilon_r(t)-h^4\norm{K(t)}}\le R_5(t,h),\qquad
 \lim_{r\to\infty}r^4\epsilon_r(t)=t^4\norm{K(t)}.
\end{equation}
If $K(t)=0$, then $\epsilon_r(t)=O(r^{-6})$.
\end{theorem}

At fixed time, accumulating one-step norms gives $t\ell$ in place of
the exact coefficient $\norm{K(t)}$ of $h^4$. Since $\norm{K(t)}\le t\ell$,
transport can reduce the leading term. At a finite step size the
additional remainder can offset that reduction.

To see how the energy gaps enter the coefficient, write
$H=\sum_a E_a P_a$ over the distinct energies of the physical
Hamiltonian. Each $P_a$ projects onto the full eigenspace, including
any degeneracy. Continuous and discrete transport act on each block
by scalar multiplication (\cref{sec:global}).
\[
 \begin{aligned}
 P_a K(t)P_b&=\left(\int_0^t e^{is(E_a-E_b)}\,ds\right)P_a L_5P_b,\\
 P_a K^d_{t,h}(B)P_b&=
 \left(h\sum_{k=0}^{r-1}e^{i(k+1/2)h(E_a-E_b)}\right)P_a BP_b.
 \end{aligned}
\]
Both kernels equal $t$ on each equal-energy block. Contractivity of
block-diagonal compression then gives
$t\norm{\sum_a P_aL_5P_a}\le\norm{K(t)}$. A nonzero energy-diagonal component rules
out complete leading-order cancellation at positive time. On distinct
energy blocks the continuous phase integral can vanish, and the
discrete sum can exhibit step-size-dependent resonances. Keeping its
finite-sum value at a resonance avoids a division by zero in numerical
evaluation.
For the Pauli pair $A=\sigma_x$, $B=\sigma_z$, write the vector of
$L_5$ in the skew-Pauli basis as the sum of a component $p_\parallel$
along $(1,0,1)/\sqrt{2}$ and a perpendicular component of signed length
$p_\perp$. The first component accumulates linearly. The second rotates
at angular frequency $2\sqrt{2}$, which gives
\[
 \norm{K(t)}^2=t^2p_\parallel^2+
      \tfrac12\sin^2(\sqrt{2}t)\,p_\perp^2 .
\]
At $t=\pi/\sqrt{2}$ the transverse contribution vanishes, while the
longitudinal component survives. The resulting leading coefficient
is about one third of $t\ell$. The exact coefficients and
a rational finite-step certificate are given in \cref{sec:examples}.

\section{From a centered defect to a global estimate}
\label{sec:method}
Both the integral and finite-sum local remainders of \citet{casas2026}
can be used for the Suzuki half-product. Their propagation across steps
requires only a bound on the one-step error and a bound on the error of
its centered approximation.

\subsection{Accumulation in the interaction picture}
\label{sec:global-transfer}
\begin{proposition}[Centered local-to-global estimate]
\label{prop:centered-global-transfer}
Let $G$ be skew-adjoint, $S(h)$ unitary, $h\ge0$, and $r\ge1$ an
integer. Suppose a matrix $B_h$ and numbers $\eta,\delta\ge0$ satisfy
\begin{equation}
 \norm{e^{-hG/2}(S(h)-e^{hG})e^{-hG/2}-B_h}\le\eta,
 \qquad \norm{S(h)-e^{hG}}\le\delta.
\end{equation}
Then
\begin{equation}
 \left\|e^{-rhG}S(h)^r-I-
 \sum_{k=0}^{r-1}e^{-(k+1/2)hG}B_he^{(k+1/2)hG}\right\|
 \le r\eta+\binom r2\delta^2.
 \label{eq:generic-global-transfer}
\end{equation}
\end{proposition}
\begin{proof}[Proof of \cref{prop:centered-global-transfer}]
Set $D_h=e^{-hG}S(h)-I$ and $D_k=e^{-khG}D_he^{khG}$. The interaction-picture product factors as
\begin{equation}
 e^{-rhG}S(h)^r=(I+D_{r-1})\cdots(I+D_0).
 \label{eq:interaction}
\end{equation}
Conjugating the centered hypothesis gives
$D_k=e^{-(k+1/2)hG}B_he^{(k+1/2)hG}+E_k$, with $\norm{E_k}\le\eta$.
Each $I+D_k$ is unitary and $\norm{D_k}\le\delta$.
For $P_k=(I+D_{k-1})\cdots(I+D_0)$ and $P_0=I$,
telescoping gives $\norm{P_k-I}\le k\delta$, while
\begin{equation}
 P_r-I-\sum_{k=0}^{r-1}D_k
 =\sum_{k=0}^{r-1}D_k(P_k-I).
\end{equation}
Since $\norm{D_k(P_k-I)}\le k\delta^2$, summing over $k$ bounds all
nonlinear terms by $\binom r2\delta^2$. Adding the $r$ local remainders
gives the result for any size of $r\delta$.
\end{proof}

The choice $B_h=0$ discards the orientation of every increment.
For the Suzuki product we take $B_h=h^5L_5+h^7N_7$. We now compute
these coefficients and bound the local remainder for finite $h$.

\subsection{Identifying the centered coefficients}
Split the central exponential to write $S(s)=U(s)U(-s)^{-1}$, where
$U$ is the six-factor half-product in \cref{eq:half}. The symmetric
defect $S'-(GS+SG)/2$ factors into unitary outer factors and the
even operator $T(s)=F(s)+F(-s)$ of \cref{eq:even-defect}.
Its $s^4$ and $s^6$ coefficients are $5L_5$ and $7M_7$, respectively.
They can be computed by multiplying noncommutative words
with coefficients in $\field$. In particular, the degree-five
coefficient of the actual step is the polynomial in \cref{eq:l5}.
The bounds on the symmetric defect follow from the local
conjugation remainder of Casas and Murua after the substitution in
\cref{eq:cm-identification}.

Conjugation by the centered half-product contributes its own
degree-two coefficient $\Omega_2$ (\cref{eq:omega2,eq:transport-jet}).
Integrating the resulting degree-six term in the centered
defect gives
\[
 N_7=M_7+\tfrac57[\Omega_2,L_5]
     =L_7+\tfrac1{24}[G,[G,L_5]].
\]
The second equality also follows by centering the term linear in the
error of $\log S(h)$. The second moment of the uniform measure on
$[-1/2,1/2]$, divided by $2!$, is $1/24$.

The nonnegative constants in \cref{eq:main-eta9} control this same
operation at finite $h$. In particular $D_9=C_9+5R_4/9+7R_2/9$,
where $R_4$ bounds the fourth-degree tail in the transport of $L_5$
and $R_2$ bounds the second-degree tail for $M_7$.
Integration of the terms involving a further copy of the local error
produces the higher powers of $h$ in $\eta_9$.
\Cref{sec:septic} proves the two local hypotheses with
$(\eta,\delta)=(\eta_9,\delta_9)$. Substitution into
\cref{prop:centered-global-transfer} proves \cref{eq:main-septic}.
At fixed $t$, $r\eta_9=O(h^8)$ and $\binom r2\delta_9^2=O(h^8)$.
A midpoint quadrature estimate is needed only when passing to the
continuous $K(t)$ in \cref{eq:main-global}.

\subsection{Comparison with accumulated local bounds}
The integral defect remainder of Casas--Murua also fits this centered
proof. Put $c=C_9^{\mathrm{CM}}$ as defined in \cref{eq:cm-constants},
and retain the transport constants $R_4,R_2$ from \cref{eq:d9}.
Define $D_9^{\mathrm{CM}}=c+5R_4/9+7R_2/9$ and obtain
$\delta_9^{\mathrm{CM}},\eta_9^{\mathrm{CM}}$ from
\cref{eq:main-delta9,eq:main-eta9} by replacing every occurrence of
$C_9,D_9$ with $c,D_9^{\mathrm{CM}}$.

Lean's propagation lemma accepts any pair of proved local bounds.
Its Suzuki instance in \cref{thm:main-septic} uses the finite-sum
constants $C_9,D_9$. For the integral constant $C_9^{\rm CM}$, the
appendices give analytic proofs of the local estimates, the comparison
of global remainders, and the threshold for strict improvement.

\begin{corollary}[Casas--Murua integral specialization of the global bound]
\label{thm:main-cm-global}
For every $t>0$ and integer $r\ge1$, let $h=t/r$ and
\begin{equation}
 R_7^{\mathrm{CM}}(t,h)=r\eta_9^{\mathrm{CM}}(h)
               +\binom r2\delta_9^{\mathrm{CM}}(h)^2.
 \label{eq:main-cm-remainder}
\end{equation}
With the same leading operator $\Lambda_{t,h}$ as above,
\begin{align}
 \norm{e^{-tG}S(h)^r-I-\Lambda_{t,h}}&\le R_7^{\mathrm{CM}}(t,h),
       \label{eq:main-cm-global}\\
 \abs{\epsilon_r(t)-\norm{\Lambda_{t,h}}}&\le R_7^{\mathrm{CM}}(t,h),
       \qquad 0\le R_7^{\mathrm{CM}}(t,h)\le R_7(t,h).
\end{align}
At fixed input and time, $R_7^{\mathrm{CM}}(t,h)=O(h^8)$.
\end{corollary}
\Cref{sec:cm-global} applies the centered estimate with the Casas--Murua
defect constant and then uses \cref{prop:centered-global-transfer}.

To state a sufficient advantage condition, define
\begin{equation}
 g(t)=t\ell-\norm{K(t)},\qquad
 J_2=\norm{[G,[G,L_5]]},\qquad \nu=\norm{N_7}.
 \label{eq:main-cm-gap}
\end{equation}

\begin{corollary}[Improvement over the accumulated Casas--Murua local bound]
\label{cor:main-cm-advantage}
Fix $t>0$. If $g(t)>0$ and $r\ge1$, $h=t/r$, satisfies
\begin{equation}
 th^2(J_2/24+\nu)+\frac{R_7^{\mathrm{CM}}(t,h)}{h^4}<g(t),
 \label{eq:main-cm-advantage}
\end{equation}
then
\begin{equation}
 \epsilon_r(t)\le\norm{\Lambda_{t,h}}+R_7^{\mathrm{CM}}(t,h)
       <th^4\ell\le r\delta_9^{\mathrm{CM}}(h).
\end{equation}
For every fixed input and time with $g(t)>0$, this condition holds
for all sufficiently large $r$.
\end{corollary}
The rightmost expression is the $q=8$ Casas--Murua local bound accumulated over
$r$ steps. An explicit sufficient threshold $r_*$, valid for all
$r\ge r_*$, and the proof are given in \cref{sec:cm-advantage}.
Both global bounds hold for every positive step count. The condition $g(t)>0$ guarantees strict improvement for all
sufficiently large $r$.

\section{What Lean checks}\label{sec:formalization}
Both estimates use the calculus of ordered products. The general theorem
takes local order as a hypothesis (\cref{thm:general-steps}). For the
eleven-factor Suzuki product, fourth-order accuracy alone leaves the
quintic and seventh-degree coefficients undetermined. To prove
\cref{thm:main-septic}, we identify both coefficients and derive explicit
expressions for the two local remainder bounds.

\begin{table}[htbp]\centering\small
\caption{Inputs and conclusions of three formal interfaces. Full names
and source locations appear in \cref{app:childs,app:lean}.}
\label{tab:interfaces}
\begin{tabular}{@{}p{.23\linewidth}p{.34\linewidth}p{.34\linewidth}@{}}
\toprule
Interface & Input & Conclusion\\\midrule
General staged product &
Signed real coefficients, stage permutations, local order, skew-adjoint summands &
Explicit commutator bound and rounded integer threshold in \cref{thm:general-steps}\\[3pt]
Centered propagation &
Contractive step, contractive exponential group, candidate centered operator and two local bounds &
Transported sum with remainder $r\eta+\binom r2\delta^2$ in \cref{eq:generic-global-transfer}\\[3pt]
Concrete Suzuki step &
Standard parameter and a Hermitian pair of complex matrices of arbitrary size &
Local finite-sum bound, quintic integral estimate and refined global operator enclosure in spectral norm\\
\bottomrule\end{tabular}
\end{table}

\paragraph{From order to a sufficient step count.}
The namespace \texttt{TrotterError} uses \texttt{Fin} for finite
indices and a reversed list product for evaluation. Its three error
identities follow from the differential equation of this ordered
product. Lean proves the four local order conditions equivalent near zero.
The resulting error bound holds at every nonnegative time. A separate
counting inequality relates the ordered commutator sum to the sum over
summands in \cref{eq:alpha-general}. This identifies the contribution
of stage placements to $\Upsilon^{p+1}$ before integration and the
final rounding of $r$.

\paragraph{From words to operators.}
The binary formulas live principally in \texttt{FQFP.BCH}. A
derivative identity for finite exponential products connects their
noncommutative coefficients to the Taylor derivatives of the actual
step. For degree seven, three certificates identify $M_7$, $N_7$
and $L_7$ with their respective 32-term right-nested expressions.
The expanded equalities cover all $2^7=128$ word positions. In particular,
\path{FQFP.BCH.suzukiCenteredSepticTerm_eq_bch} proves
$N_7=L_7+[G,[G,L_5]]/24$.

In \path{InteractionPicture.lean}, the propagation lemma has the
following statement in the ambient algebra. Its argument \texttt{B}
specifies the one-step operator whose conjugates are summed before
taking the norm.
\begin{lstlisting}[
  basicstyle=\small\ttfamily, columns=fullflexible, keepspaces=true,
  morekeywords={theorem}, keywordstyle=\bfseries\color{MidnightBlue},
  frame=leftline, rulecolor=\color{MidnightBlue!45}, framesep=6pt,
  xleftmargin=10pt, aboveskip=5pt, belowskip=4pt,
  literate={∀}{{$\forall$}}1 {ℝ}{{$\mathbb R$}}1 {ℕ}{{$\mathbb N$}}1
           {‖}{{$\Vert$}}1 {•}{{$\bullet$}}1 {≤}{{$\le$}}1
           {η}{{$\eta$}}1 {δ}{{$\delta$}}1 {∑}{{$\sum$}}1 {∈}{{$\in$}}1
]
theorem norm_interactionProduct_sub_sum_conj_le
    (G S B : A)
    (hG : ∀ t : ℝ, ‖exp (t • G)‖ ≤ 1) (hS : ‖S‖ ≤ 1)
    (h : ℝ) {η δ : ℝ}
    (hlocal : ‖exp ((-h / 2) • G) * (S - exp (h • G)) *
                exp ((-h / 2) • G) - B‖ ≤ η)
    (hδ : ‖S - exp (h • G)‖ ≤ δ) (r : ℕ) :
    ‖exp (-((r : ℝ) * h) • G) * S ^ r - 1 -
      ∑ k ∈ range r, expSMulConj (-G) B (((k : ℝ) + 1 / 2) * h)‖ ≤
        (r : ℝ) * η + (r : ℝ) * ((r : ℝ) - 1) / 2 * δ ^ 2
\end{lstlisting}

\paragraph{Scope of the declarations.}
The general skew-adjoint commutator bound uses a nontrivial complete
C*-ring with normalized unit and a compatible real scalar action.
The zero-dimensional matrix case of the inequality has zero error.
For \cref{prop:centered-global-transfer}, the Lean statement above admits
any real $h$ and natural $r$ under its contraction and local error
hypotheses. The binary Suzuki development derives the exponential
contractions for $X,Y,G$ from skew-adjointness. On setting $X=-iA$
and $Y=-iB$, \path{formalization/MatrixBounds.lean} checks
\cref{eq:main-local,eq:main-global,eq:main-septic} in the spectral
norm for Hermitian pairs of complex $d\times d$ matrices of arbitrary $d$.

Lean proves the quintic comparison $\ell\le B_J/8$, the finite-sum
operator bound and its scalar enclosure with $R_7$, and the fixed-time
limit $t^4\norm{K(t)}$. The sufficient single-step condition with
decimal constant $0.97$ uses a separate rational-interval calculation.
The integral defect constant $C_9^{\rm CM}$, the conclusions about
$R_7^{\rm CM}$ and its strict-advantage threshold, the energy-block
formulas, and the three-generator extension have analytic proofs in
the appendices. Of the 105 many-body instances, 35 use the verified
binary formula and 70 use the three-generator extension. All reported
many-body norm evaluations and step counts use floating-point arithmetic
(\cref{app:childs,app:lean,app:repro}).

\section{Numerical consequences for simulation step counts}
\label{sec:numerical-summary}
For each Hamiltonian and tolerance $\varepsilon$, an integer $r$ suffices
if $\norm{\Lambda_{t,h}}+R_7(t,h)\le\varepsilon$. Testing this bound
requires evaluating the norm of the transported sum at each trial $r$.
The data comprise three disordered Heisenberg configurations,
seven sizes $n=4,\ldots,10$ and five saved fields per configuration
and size. The time is $t=n$ and the spectral-norm tolerance is
$10^{-3}$. The nearest-neighbor model uses an even--odd splitting
into two terms. Each local commutator baseline is evaluated on the
same matrices as the transported error. \Cref{sec:experiments} specifies
the Hamiltonians, symmetry blocks and step-count searches.

\begin{table}[htbp]\centering\small
\caption{Mean step counts on the same five ten-site nearest-neighbor
instances at $t=10$ and tolerance $10^{-3}$. The second and third rows
use identical signed coefficients and an identical grouped local tail.
The penultimate row evaluates the finite-sum global remainder $R_7$.}
\label{tab:main-steps}
\begin{tabular}{@{}lr@{}}\toprule
Estimate used to select the step count & Mean steps\\\midrule
Childs et al., Proposition J.1 & 585.4\\
Grouped local tail, separate norms of six quintic terms & 247.4\\
Same tail, norm of the complete $L_5$ & 190.6\\
Accumulated Casas--Murua local bound, $q=8$ & 184.8\\
Transported $L_5$ and $N_7$ with finite-sum remainder & 129.4\\
Direct simulation, empirical crossing & 123.8\\\bottomrule
\end{tabular}
\end{table}

\paragraph{Local grouping and the complete estimators.}
The shift from $247.4$ to $190.6$ in \cref{tab:main-steps} isolates the
effect of adding the six signed quintic terms before taking their norm.
Accumulating the Casas--Murua local bounds at $q=4,6,8$ gives $312.4$,
$187.8$ and $184.8$ steps, respectively. The finite-sum septic local
estimate also gives $184.8$ on these samples. The transported finite-sum
estimate gives $129.4$, which is $4.52\%$ above the empirical mean.
The $q=8$ local estimate uses the defect coefficient $M_7$. The
transported estimate uses the centered coefficient $N_7$ and a
different remainder bound. We evaluated the $q=8$ local comparison
on all 35 binary instances (\cref{tab:cm-steps,tab:leading-ablation}).

\paragraph{Holding the polynomial and remainder fixed.}
To measure cross-step cancellation alone, fix $B_h=h^5L_5+h^7N_7$
and the finite-sum remainder $R_7$ of \cref{thm:main-septic}.
Unitary conjugation preserves each step's norm, so taking the norms
separately gives $U_{\rm sep}=r\norm{B_h}+R_7$. Adding the same
conjugates first gives $U_{\rm tr}=\norm{\Lambda_{t,h}}+R_7$.
The stored data contain $\ell=\norm{L_5}$ and $\nu=\norm{N_7}$.
The triangle and reverse triangle inequalities therefore give
\[
 r\,|h^5\ell-h^7\nu|+R_7\ \le\ U_{\rm sep}\
      \le\ r(h^5\ell+h^7\nu)+R_7 .
\]
For each field we test the two envelopes at successive integers
$r=1,2,\ldots$. Their first crossings bracket the first crossing of
$U_{\rm sep}$ even if the bounds vary nonmonotonically. For $U_{\rm tr}$
we use the stored count where the bound passes at $r$ and fails at $r-1$.
This adjacent check establishes feasibility at the reported count.

\begin{table}[htbp]\centering\small
\caption{Controlled comparison on the five nearest-neighbor fields at
$n=t=10$ and tolerance $10^{-3}$. Both columns use the same $B_h$ and
$R_7$. Ranges reflect bounds on the joint norm calculated from saved
floating-point inputs.}
\label{tab:controlled-transport}
\begin{tabular}{@{}lrr@{}}\toprule
Field & Separate norms, first crossing & Transported sum, saved crossing\\\midrule
0 & 186--187 & 130\\
1 & 184--185 & 129\\
2 & 182--183 & 128\\
3 & 181 & 129\\
4 & 190 & 131\\
\midrule
Mean & 184.6--185.2 & 129.4\\\bottomrule
\end{tabular}

\end{table}

On these fields the mean first crossing for separate norms lies
between $184.6$ and $185.2$ steps. The transported sum meets tolerance at
a mean of $129.4$ steps, a reduction of $29.9$--$30.1\%$ in the ratio
of means. The ranges in \cref{tab:controlled-transport} come from upper and lower
estimates of $\norm{B_h}$ using saved floating-point values. Their
width measures neither sampling uncertainty nor floating-point error.
The bundled \path{compare_transport.py} reproduces the table.

\citet[Fig.~9]{liu2026} also evaluate fourth-order Suzuki steps for
five random-field Heisenberg chains at tolerance $10^{-3}$, including
the even--odd ordering. They accumulate a local-region bound up to
256 qubits at $t=n+1$. Here $t=n$ and the transported norm uses dense
symmetry blocks. All numerical ratios above compare methods on the
same five fields and at the same time within this study.

\begin{landscape}
\begin{figure}[p]\centering
\includegraphics[width=\linewidth]{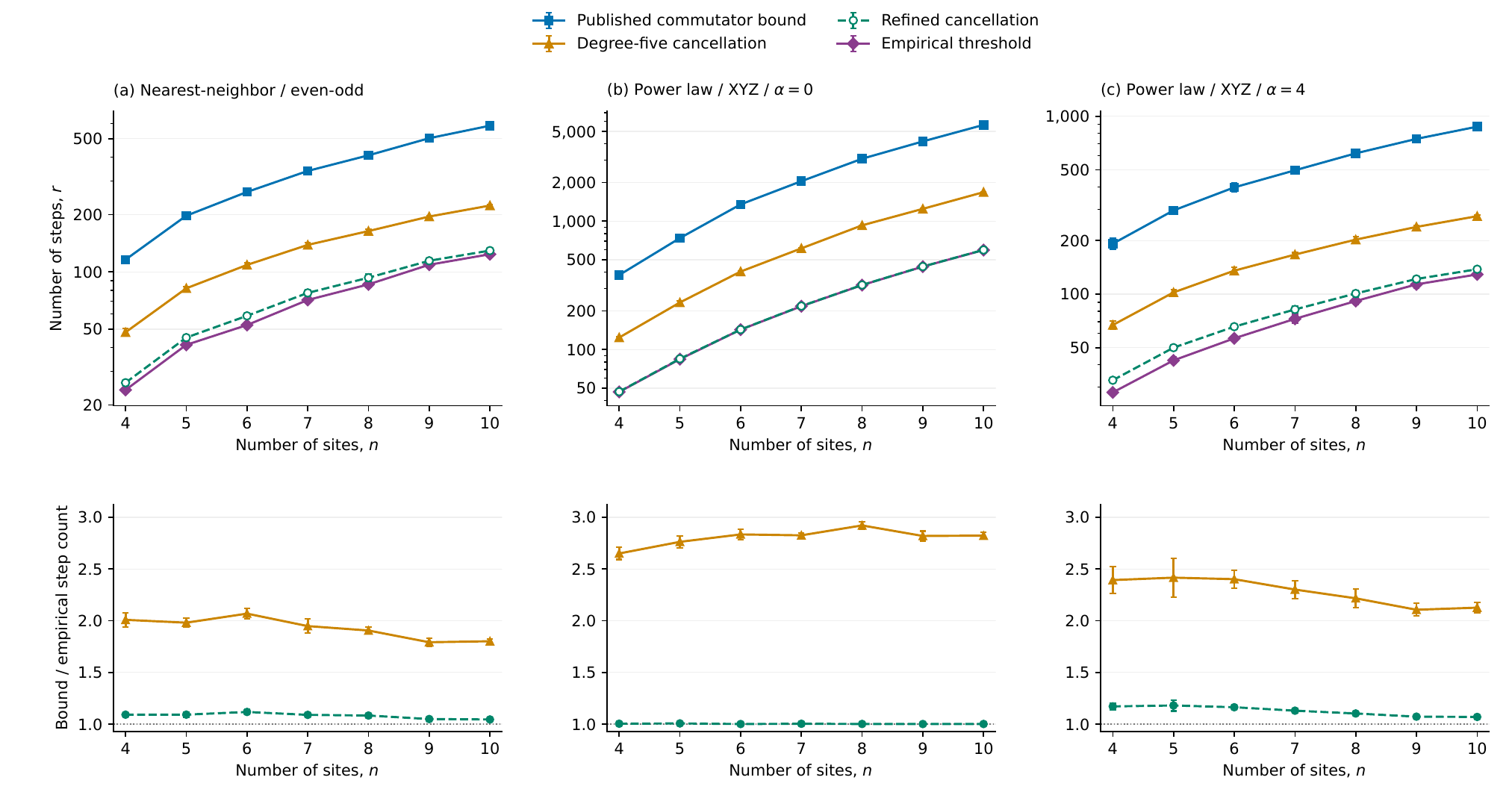}
\caption{Step counts for 105 disordered Heisenberg instances at $t=n$
and tolerance $10^{-3}$. Top: the published commutator bound, the
degree-five cancellation bound, the refined bound in
\cref{eq:septic-norm}, and direct simulation. Bottom: the mean of
the per-instance ratios of bound-based to empirical step counts.
Points and error bars show means and sample standard deviations over
five fields. Every term of \cref{eq:eta9,eq:septic-norm} is included.}
\label{fig:septic}
\end{figure}
\end{landscape}

\paragraph{Dependence on the splitting.}
The other 70 instances use a three-term XYZ splitting of power-law
models at exponents $\alpha=0,4$. The required 21-factor instance
is derived in \cref{sec:three}. At $n=10$ and $\alpha=0$ the Childs
et al.\ Appendix J.2 bound gives $5623.0$ steps. Retaining the
septic term locally gives $608.4$, and transporting the two
polynomials gives $597.2$, compared with the empirical $596.2$.
Most of the gain here comes from the local estimate. At $\alpha=4$ the corresponding means are $873.4$, $205.0$, $138.0$
and $129.0$, so transport accounts for a larger share of the reduction. \Cref{fig:septic}
and \cref{app:refined-steps} show all sizes. Every reported empirical
crossing passes at $r$ and fails at $r-1$. Coarse-step errors can
oscillate, so these adjacent values identify a local crossing in the saved
floating-point search. Interval bounds for the separate Pauli example
are given in \cref{sec:examples}.

\section{Scope and conclusions}\label{sec:limitations}
For staged formulas satisfying \cref{thm:general-steps}, the local
order and a sum of nested commutator norms determine a sufficient step
count. Fixing the Suzuki formula and the Hamiltonian lets us retain
two kinds of cancellation. The six signed terms of $L_5$ can cancel
within a step. Across steps, the exact dynamics conjugates both $L_5$
and $N_7$ before their contributions are added. The resulting sum
bounds the finite-step error once the explicit remainder is included.
At leading order, the part of $L_5$ commuting with $H$ accumulates
linearly, while transverse components can average out over the chosen
time.

Which bound is practical depends on what one can compute. Local norm
sums can be estimated from the Hamiltonian summands. Evaluating
$\norm{\Lambda_{t,h}}$ also requires the transported operators, for
which our experiments use dense symmetry blocks. A comparison on larger
systems would require an efficient approximation with controlled error
for these operators or their norms.

The estimates apply to bounded generators. Unbounded Hamiltonians
require additional results on domains and evolution. For splittings
with more terms or for higher Suzuki orders, the same propagation lemma
applies once the coefficient identities and local remainders have been
established.

\begingroup
\small
\setlength{\bibsep}{3pt}
\bibliographystyle{plainnat}
\bibliography{references}
\endgroup
\clearpage
\appendix
\crefalias{section}{appendix}
\section{The exact degree-five Hall polynomial}
\label{sec:hall}

\subsection{Basis and coefficient field}
Use the following classical Hall elements:
\begin{equation}
\begin{aligned}
 E_1&=X,& E_2&=Y,& E_3&=[Y,X],\\
 E_4&=[E_3,X],&E_5&=[E_3,Y],\\
 E_6&=[E_4,X],&E_7&=[E_4,Y],&E_8&=[E_5,Y],\\
 E_9&=[E_6,X],&E_{10}&=[E_6,Y],&E_{11}&=[E_7,Y],\\
 E_{12}&=[E_8,Y],&E_{13}&=[E_4,E_3],&E_{14}&=[E_5,E_3].
\end{aligned}\label{eq:hall-basis}
\end{equation}
This recursive definition fixes the Hall-basis signs throughout.
In particular, the final brackets in $E_{13}$ and $E_{14}$ use $[Y,X]$.

The algebraic number $u$ satisfies
\begin{equation}
 60u^3-48u^2+12u-1=0,\qquad
 0.414490771794375<u<0.414490771794376.
 \label{eq:field}
\end{equation}
Consequently every exact coefficient is stored as $q_0+q_1u+q_2u^2$,
with $q_j\in\mathbb Q$. Multiplication reduces powers using
$u^3=\frac45u^2-\frac15u+\frac1{60}$.

\begin{proposition}[Exact modified generator]
In the formal free associative algebra, or as a local matrix expansion,
\begin{equation}
 \log S(h)=hG+h^5L_5+O(h^7),\qquad
 L_5=\sum_{j=9}^{14}\ell_jE_j,
 \label{eq:l5}
\end{equation}
where the coefficients are given in \cref{tab:hall}.
\end{proposition}

\par
\begin{table}[htbp]\centering
\caption{Exact Hall coefficients for the specified Suzuki composition.
The common denominator in the middle column is $144000$.}
\label{tab:hall}
\begin{tabular}{cl r}\toprule
Element & $144000\ell_j$ & Approximate $\ell_j$\\\midrule
$E_9$ & $6-52u-127u^2$ & $-0.000259530905007$\\
$E_{10}$ & $24-208u-508u^2$ & $-0.001038123620026$\\
$E_{11}$ & $36-312u-512u^2$ & $-0.001258917027431$\\
$E_{12}$ & $24-208u-8u^2$ & $-0.000441586814809$\\
$E_{13}$ & $36-312u-12u^2$ & $-0.000662380222213$\\
$E_{14}$ & $72-624u+976u^2$ & $-0.000131686833991$\\\bottomrule
\end{tabular}
\end{table}
\par

\subsection{A derivation by graded BCH composition}
For one Strang step write
\begin{equation}
 \log S_2(h)=hG+h^3C_3+h^5D_5+O(h^7),
\end{equation}
where finite multiplication and logarithm expansion give
\begin{align}
 C_3&=-\frac{E_4}{24}-\frac{E_5}{12},\label{eq:c3}\\
 D_5&=\frac{7E_9+28E_{10}+32E_{11}+8E_{12}+12E_{13}-16E_{14}}{5760}.
 \label{eq:d5}
\end{align}
For the five-stage composition we obtain
\begin{equation}
 \boxed{L_5=\kappa D_5+\lambda[G,[G,C_3]],}\label{eq:l5-decomp}
\end{equation}
with
\begin{align}
 \kappa&=4u^5+v^5=\frac{-6+52u-123u^2}{75},\\
 \lambda&=\frac{-3+26u-24u^2}{900}.
\end{align}
The second term comes from commutators between Strang stages. To
compute it, write the logarithm of a partial product through degree
five as
\begin{equation}
 ahG+bh^3C_3+ch^4[G,C_3]
       +h^5\bigl(dD_5+e[G,[G,C_3]]\bigr).
\end{equation}
Append $S_2(qh)$ on the right and put $\Delta=aq^3-bq$. The BCH formula
through the required degree yields
\begin{equation}
\begin{aligned}
 a'&=a+q,& b'&=b+q^3,&c'&=c+\Delta/2,\\
 d'&=d+q^5,&e'&=e-cq/2+(a-q)\Delta/12.
\end{aligned}\label{eq:recurrence}
\end{equation}
For example, $\tfrac12[A,B]$ contributes $\Delta[G,C_3]/2$ at degree four
and $-cq[G,[G,C_3]]/2$ at degree five. The two double-bracket BCH terms
contribute $(a-q)\Delta[G,[G,C_3]]/12$.
Starting from zero and successively setting $q=u,u,v,u,u$ gives
$(a,b,c,d,e)=(1,0,0,\kappa,\lambda)$.
Expanding the double bracket in \cref{eq:l5-decomp} in the basis
\cref{eq:hall-basis} gives \cref{tab:hall}.

As an independent check, the coefficient script multiplies all eleven
exponential series through degree six and computes the formal logarithm
of the product. It verifies every one of
the $2^5=32$ degree-five words, as well as the disappearance of degrees
two, three, four, and six in the logarithm.

\subsection{A right-nested representation for comparisons}
Set $R_{abc}=\ad_a\ad_b\ad_c[Y,X]$. The Jacobi identity implies
\begin{equation}
 R_{XXY}=R_{XYX},\qquad R_{YXY}=R_{YYX}.
\end{equation}
For instance their first difference is
$\ad_X[\ad_X,\ad_Y][Y,X]=\ad_X\ad_{[X,Y]}[Y,X]=0$.
It follows that
\begin{equation}
 L_5=\sum_{w\in\mathcal W}d_wR_w,\qquad
 \mathcal W=\{XXX,XXY,XYY,YXX,YXY,YYY\},
 \label{eq:right-l5}
\end{equation}
with the positive coefficients in \cref{tab:right}. Although the scalar coefficients are positive in this representation,
the operators can cancel in the sum.

\par
\begin{table}[htbp]\centering
\caption{Right-nested degree-five coefficients and the combined displayed
coefficients $b_w$ from the published two-term proposition.}
\label{tab:right}
\begin{tabular}{clrr}\toprule
$w$ & $144000d_w$ & $d_w$ (approx.) & $b_w$\\\midrule
$XXX$ & $-6+52u+127u^2$ & $0.000259531$ & $0.0047$\\
$XXY$ & $-36+312u+12u^2$ & $0.000662380$ & $0.0103$\\
$XYY$ & $-72+624u-976u^2$ & $0.000131687$ & $0.0074$\\
$YXX$ & $12-104u+496u^2$ & $0.000375743$ & $0.0097$\\
$YXY$ & $36-312u+1488u^2$ & $0.001127230$ & $0.0270$\\
$YYY$ & $-24+208u+8u^2$ & $0.000441587$ & $0.0284$\\\bottomrule
\end{tabular}
\end{table}
\par

\section{Suzuki factorization and the original local constant}
\label{sec:local}

\subsection{A finite conjugation estimate}
Commutator words run from outermost to innermost;
generator lists in $\Acal_p$ run from innermost to outermost.
For anti-Hermitian matrices $Z_1,\ldots,Z_s$ and any matrix $B$, define,
for an integer $p\ge1$,
\begin{equation}
 \Acal_p(Z_1,\ldots,Z_s;B)=
 \sum_{q_1+\cdots+q_s=p}\frac{p!}{q_1!\cdots q_s!}
 \norm{\ad_{Z_s}^{q_s}\cdots\ad_{Z_1}^{q_1}B}.
 \label{eq:ap}
\end{equation}
The empty-list value is zero. The associated product of conjugations is
\begin{equation*}
 \Phi(x)=e^{x\ad_{Z_s}}\cdots e^{x\ad_{Z_1}}B.
\end{equation*}
The finite conjugation expansion in~\cite{childs} states that its Taylor
polynomial through degree $p-1$ has remainder bounded by
\begin{equation}
 \norm{\Phi(x)-\sum_{k=0}^{p-1}x^k\Phi_k}
 \le \frac{|x|^p}{p!}\Acal_p(Z_1,\ldots,Z_s;B).
 \label{eq:finite-conj}
\end{equation}
Here $\Phi_k$ is the degree-$k$ coefficient of the exponential
product. Unitarity makes the estimate valid for every real $x$.

\subsection{Symmetric factorization and defect}
For the two-term formula set
\begin{equation}
 (Z_1,\ldots,Z_6)=
 \left(\frac u2X,uY,uX,uY,\frac{1-3u}{2}X,\frac v2Y\right).
 \label{eq:half}
\end{equation}
Then $\sum_jZ_j=G/2$ and
\begin{equation}
 U(s)=e^{sZ_1}\cdots e^{sZ_6},\qquad V(s)=U(-s)^{-1},
 \qquad S(s)=U(s)V(s).
\end{equation}
Splitting the central exponential in half explains the six factors in $U$
despite the eleven merged factors in $S$.
Define the symmetric defect
\begin{equation}
 \mathcal D(s)=S'(s)-\tfrac12\bigl(GS(s)+S(s)G\bigr).
 \label{eq:defect}
\end{equation}
Writing $\mathcal D=UTV$, introduce
\begin{equation}
 P(s)=U(s)^{-1}U'(s),\quad Q(s)=U(s)^{-1}GU(s),\quad
 F(s)=P(s)-\tfrac12Q(s).
\end{equation}
Direct differentiation of $V(s)=U(-s)^{-1}$ gives the exact identity
\begin{equation}
 \boxed{T(s)=F(s)+F(-s).}\label{eq:even-defect}
\end{equation}
In particular, $T$ is even. The conjugations occurring in $F$ are
\begin{align}
 P(s)&=\sum_{j=1}^6 e^{-s\ad_{Z_6}}\cdots
                   e^{-s\ad_{Z_{j+1}}}Z_j,\\
 Q(s)&=e^{-s\ad_{Z_6}}\cdots e^{-s\ad_{Z_1}}G.
\end{align}

\begin{definition}[Seventh-order commutator constant]
\begin{equation}
 C_7=\frac{1}{7!}\left(
 2\sum_{j=1}^6\Acal_6(Z_{j+1},\ldots,Z_6;Z_j)
 +\sum_{B\in\{X,Y\}}\Acal_6(Z_1,\ldots,Z_6;B)\right).
 \label{eq:c7}
\end{equation}
\end{definition}
This is a finite, nonnegative sum of norms of degree-seven commutators.
Taking absolute values of scalar stage coefficients and orienting the
innermost bracket as $[Y,X]$ rewrites it as
\begin{equation}
 C_7=\sum_{w\in\{X,Y\}^5}\rho_w
       \norm{\ad_{w_1}\cdots\ad_{w_5}[Y,X]},\qquad \rho_w\ge0.
 \label{eq:rho}
\end{equation}
The coefficient data give exact algebraic values for all 32 weights.
They bound the remainder. The seventh-degree BCH polynomial has its
own coefficients.

\begin{corollary}[Local bound from the cited conjugation estimate]
\label{thm:local}
For all real $h$,
\begin{equation}
 \boxed{\norm{S(h)-e^{hG}}\le |h|^5\norm{L_5}+|h|^7C_7.}
 \label{eq:local}
\end{equation}
Moreover, $\lim_{h\to0^+}h^{-5}\norm{S(h)-e^{hG}}=\norm{L_5}$.
\end{corollary}
\begin{proof}
Order four gives $T(s)=O(s^4)$. The degree-five local error is $s^5L_5$,
so differentiating the finite jet of $S$ in \cref{eq:defect} gives
$[s^4]T=5L_5$. Apply \cref{eq:finite-conj} with $p=6$ to the
conjugations in $F(s)$ and $F(-s)$. Odd polynomial terms cancel by
\cref{eq:even-defect}, and the constant and quadratic terms vanish by
order four. Splitting $G=X+Y$ only in the remainder yields
\begin{equation}
 \norm{T(s)-5s^4L_5}\le7C_7|s|^6.
 \label{eq:central-rem}
\end{equation}
The symmetric variation-of-constants formula is
\begin{equation}
 S(h)-e^{hG}=\int_0^h
 e^{(h-s)G/2}\mathcal D(s)e^{(h-s)G/2}\,ds.
 \label{eq:duhamel}
\end{equation}
For $h\ge0$, integrating the defect bound in this symmetric Duhamel
identity gives the inequality \cite{auzinger}. Symmetry extends it to
negative $h$. The expansion $S(h)-e^{hG}=h^5L_5+O(h^6)$ gives the limit.
The grouped and integral versions are the Casas--Murua specializations in
\cref{sec:cm}; no equality between the original $C_7$ and the grouped
constant is assumed.
\end{proof}

\subsection{Comparison with the published commutator bound}
Proposition J.1 of~\cite{childs}, after the Jacobi reductions in
\cref{sec:hall}, has the form
\begin{equation}
 \norm{S(h)-e^{hG}}\le |h|^5\bp,\qquad
 \bp=\sum_{w\in\mathcal W}b_w\norm{R_w}.
 \label{eq:paper-local}
\end{equation}
The $b_w$ in \cref{tab:right} are the displayed decimal coefficients,
treated as exact rationals. In the lexicographic order $XXX,XXY,\ldots,YYY$,
let $c_i$ be the signed quintic coefficients before combining equal
commutators, using the representation with $c_{XYX}=c_{YYX}=0$,
and let $\tilde b_i$ be the corresponding eight published weights.
At the standard parameter, rational upper bounds for $10^6|c_i|$ are
$(260,663,0,132,376,1128,0,442)$.
Each is at most $10^6\tilde b_i/8$. The Jacobi equalities in \cref{sec:hall}
combine the weights $0.0057+0.0046=0.0103$ and $0.0097+0.0173=0.0270$,
giving exactly the six weights in \cref{tab:right}. The triangle inequality yields
\begin{equation}
 \ell=\norm{L_5}\le\sum_i |c_i|\norm{R_i}
       \le \frac18\sum_i\tilde b_i\norm{R_i}=\frac{\bp}{8}.
 \label{eq:quintic-eighth}
\end{equation}
For the finite-step comparison we also bound $C_7$.

Let $M=\max(\norm{X},\norm{Y})$. Removing the outermost two brackets in
\cref{eq:rho} by $\norm{[Z,B]}\le2\norm{Z}\norm{B}$ gives
\begin{equation}
 C_7\le4M^2\sum_{w\in\mathcal W}W_w\norm{R_w},\qquad W_w\ge0.
\end{equation}
Each $W_w$ is the sum of weights whose remaining three-letter tail reduces
to $w$. The six values are recorded in \cref{tab:w}.
\par
\begin{table}[htbp]\centering
\caption{Reduced seventh-order weights. The last column is a numerical
diagnostic; the sufficient condition $0.97$ is checked by rational intervals.}
\label{tab:w}
\begin{tabular}{crr}\toprule
$w$ & $W_w$ (approx.) & $\sqrt{(b_w-d_w)/(4W_w)}$\\\midrule
$XXX$ & $0.00116015258204$ & $0.978198$\\
$XXY$ & $0.00208275752566$ & $1.075562$\\
$XYY$ & $0.00156621225997$ & $1.077114$\\
$YXX$ & $0.00157889030986$ & $1.215070$\\
$YXY$ & $0.00305614150556$ & $1.454805$\\
$YYY$ & $0.00227379010405$ & $1.753279$\\\bottomrule
\end{tabular}
\end{table}
\par
In particular,
\begin{equation}
 |h|^5\norm{L_5}+|h|^7C_7
 \le |h|^5\sum_{w\in\mathcal W}(d_w+4h^2M^2W_w)\norm{R_w}.
\end{equation}
Exact interval arithmetic using \cref{eq:field} verifies all six inequalities
$b_w-d_w-4(97/100)^2W_w>0$. Thus:
\begin{corollary}
If $|h|M\le0.97$, the right-hand side of \cref{eq:local} is no larger
than the bound of Proposition J.1. The improvement is strict whenever
$|h|^5\bp>0$.
\end{corollary}
This sufficient condition is not necessary. For fixed matrices a direct
criterion is $h^2C_7<\bp-\norm{L_5}$, provided the right side is positive.
At any step one may use the minimum of the two bounds and the unitary
distance bound $2$.

\section{Casas--Murua local bounds at the Suzuki coefficients}
\label{sec:cm}
We apply the conjugation remainder and grouping argument of
\citet{casas2026} to the Suzuki half-list in \cref{eq:half} with the
coefficients given there. The integral and finite-sum forms supply
the constants used in the computations and the Lean proofs.

\subsection{Recurrence and coefficient identification}
The defect in \cref{eq:even-defect} has the recurrence
\begin{equation}
 F_0(x)=-G/2,\qquad
 F_j(x)=e^{-x\ad_{Z_j}}F_{j-1}(x)+Z_j,\qquad F=F_s.
 \label{eq:f-recursive}
\end{equation}
Use the positive-sign convention of \citet{casas2026} by setting
\begin{equation}
 \widehat F_0(x)=-G/2,\qquad
 \widehat F_j(x)=e^{x\ad_{Z_j}}\widehat F_{j-1}(x)+Z_j.
 \label{eq:cm-recurrence}
\end{equation}
This is \cite[(54), (56)]{casas2026}, with initial generator $-G/2$
followed by $Z_1,\ldots,Z_s$. Consequently
\begin{equation}
 \widehat F_j(x)=F_j(-x),\qquad
 \widehat f_{j,k}=(-1)^kf_{j,k},\qquad f_{j,k}=[x^k]F_j(x).
 \label{eq:cm-identification}
\end{equation}
For $\Theta(x)=U(-x)^{-1}e^{-xG/2}$, differentiation gives
$\Theta'\Theta^{-1}=\widehat F_s$ and
$S(h)-e^{hG}=U(h)(\Theta(h)-\Theta(-h))e^{hG/2}$.
Thus \cite[Lemma~4.6]{casas2026} applies to the same local error for $h\ge0$.
The actual fourth-order Suzuki jet gives
\begin{equation}
 f_{s,0}=f_{s,2}=0,\qquad 2f_{s,4}=5L_5,\qquad
 M_7=\tfrac27 f_{s,6},\qquad \mu=\norm{M_7}.
 \label{eq:m7}
\end{equation}
The first three identities follow by inserting the product jet from
\cref{sec:hall} into the defect; the last defines the retained
degree-seven defect coefficient. Its independent coefficient checks
are recorded in \cref{sec:septic}.

\subsection{Cited remainder and its finite-sum form}
Let $A_1,\ldots,A_s$ be anti-Hermitian and let $B,K_1,\ldots,K_s$ be
fixed matrices. Consider the affine conjugation recurrence
\begin{equation}
 B_0(x)=B,\qquad B_j(x)=e^{x\ad_{A_j}}B_{j-1}(x)+K_j.
 \label{eq:affine-recursion}
\end{equation}
Write $b_{j,k}=[x^k]B_j(x)$ for its Taylor coefficients, including
factorials. They can be computed without diagonalization:
\begin{equation}
 b_{j,k}=\sum_{q=0}^k\frac{\ad_{A_j}^{q}}{q!}b_{j-1,k-q}
                 +\mathbf1_{k=0}K_j,\quad
 b_{0,0}=B,\quad b_{0,k}=0\ (k>0).
 \label{eq:grouped-coefficients}
\end{equation}
For an integer $p\ge1$, define the grouped remainder constant
\begin{equation}
 \mathfrak B_p(A,K;B)=\sum_{j=1}^s\sum_{k=0}^{p-1}
 \frac{\norm{\ad_{A_j}^{p-k}b_{j-1,k}}}{(p-k)!}.
 \label{eq:grouped-constant}
\end{equation}

The finite-sum form in \cite[Lemma~4.3 and Remark~4.5]{casas2026} gives, for every
real $x$,
\begin{equation}
 \norm{B_s(x)-\sum_{k=0}^{p-1}x^kb_{s,k}}
 \le |x|^p\mathfrak B_p(A,K;B).
 \label{eq:grouped-bound}
\end{equation}
Fixed kicks affect only the constant coefficient, so the same cited
conjugation remainder applies; zero kicks give the pure transport case.
For the defect recurrence define
\begin{equation}
 B_p^F=\sum_{j=1}^s\sum_{k=0}^{p-1}
 \frac{\norm{\ad_{Z_j}^{p-k}f_{j-1,k}}}{(p-k)!},\qquad
 C_7^{\mathrm g}=\tfrac27B_6^F,\qquad C_9=\tfrac29B_8^F.
 \label{eq:refined-constants}
\end{equation}
For even positive $q$, the corresponding Casas--Murua integral constant is
\begin{equation}
 \widetilde B_q^F=\sum_{j=1}^s\int_0^1
 \left\|\sum_{n=1}^q\frac{\sigma^{n-1}}{(n-1)!}
 \ad_{Z_j}^n\widehat f_{j-1,q-n}\right\|\,d\sigma.
 \label{eq:cm-integral}
\end{equation}
By \cite[Lemma~4.3]{casas2026}, applied at $x$ and $-x$,
\begin{equation}
 \left\|\widehat F_s(x)-\sum_{k=0}^{q-1}x^k\widehat f_{s,k}\right\|
 \le |x|^q\widetilde B_q^F.
 \label{eq:cm-jet-remainder}
\end{equation}
Its unitary-conjugation proof applies to either sign of $x$.
Taking separate norms in \cref{eq:cm-integral} and integrating the
scalar weights gives exactly the relaxation in \cite[Remark~4.5]{casas2026}:
\begin{equation}
 \widetilde B_q^F\le
 \sum_{j=1}^s\sum_{n=1}^q
 \frac{\norm{\ad_{Z_j}^n\widehat f_{j-1,q-n}}}{n!}=B_q^F.
 \label{eq:cm-domination}
\end{equation}
Set
\begin{equation}
 C_5^{\mathrm{CM}}=\tfrac25\widetilde B_4^F,\qquad
 C_7^{\mathrm{CM}}=\tfrac27\widetilde B_6^F,\qquad
 C_9^{\mathrm{CM}}=\tfrac29\widetilde B_8^F.
 \label{eq:cm-constants}
\end{equation}

\begin{corollary}[Casas--Murua estimates at the Suzuki coefficients]
\label{lem:cm-defect}
For every real $h$,
\begin{align}
 \norm{S(h)-e^{hG}}&\le |h|^5C_5^{\mathrm{CM}},\label{eq:cm-four}\\
 \norm{S(h)-e^{hG}}&\le |h|^5\ell+|h|^7C_7^{\mathrm{CM}},\label{eq:cm-six}\\
 \norm{S(h)-e^{hG}}&\le |h|^5\ell+|h|^7\mu+|h|^9C_9^{\mathrm{CM}}.
 \label{eq:cm-eight}
\end{align}
The latter two also hold with $C_7^{\mathrm g}$ and $C_9$,
respectively:
\begin{align}
 \norm{S(h)-e^{hG}}&\le |h|^5\ell+|h|^7C_7^{\mathrm g},\label{eq:grouped-seven}\\
 \norm{S(h)-e^{hG}}&\le |h|^5\ell+|h|^7\mu+|h|^9C_9.\label{eq:local-nine}
\end{align}
For $c=C_9$ or $C_9^{\mathrm{CM}}$ and every real $x$,
\begin{equation}
 \norm{T(x)-5x^4L_5-7x^6M_7}\le9c|x|^8.
 \label{eq:defect-nine}
\end{equation}
\end{corollary}
\begin{proof}
Use \cite[Proposition~4.7 and Remark~4.8]{casas2026} at order four,
with $q=6,8$ and \cref{eq:m7}.
For the defect estimate, add the $q=8$ expansions in
\cref{eq:cm-jet-remainder} at $x$ and $-x$:
odd terms cancel and the two remainders sum to at most
$2|x|^8\widetilde B_8^F=9C_9^{\mathrm{CM}}|x|^8$.
Integrating the even defect gives the displayed local bounds for
$h\ge0$; symmetry and unitary invariance give negative steps.
The finite-sum versions follow from \cref{eq:cm-domination}.
\end{proof}
The integral constant is no larger than its finite-sum counterpart.
The saved global step counts below use \cref{eq:septic-norm}, with
finite-sum constants.

\subsection{Numerical comparison on identical inputs}
The supplemental comparison script evaluates \cref{eq:cm-integral}
by Gauss--Legendre quadrature. For a symmetry-block decomposition,
it first takes the maximum block norm at each stage and quadrature
node, then integrates and sums over stages. This order evaluates the
full-space norm in \cref{eq:cm-integral}.

\Cref{tab:cm-pauli} gives the local constants for the Pauli pair of
\cref{sec:examples}.
\begin{table}[htbp]\centering\small
\caption{Pauli local constants. Integral entries are floating-point
quadrature evaluations.}
\label{tab:cm-pauli}
\begin{tabular}{lrr}\toprule
Coefficient & Finite sum & Casas--Murua integral\\\midrule
Seventh-order tail & $0.0427241247888$ & $0.0266552702109$\\
Ninth-order tail & $0.00784875832988$ & $0.00357795701091$\\\bottomrule
\end{tabular}
\end{table}
Here $C_5^{\mathrm{CM}}\simeq0.1172427677201$ and
$\ell\simeq0.01449815432796$. The 32-, 64-, and 128-node evaluations
agree to the displayed precision.

\Cref{tab:cm-steps} compares the three accumulated local estimates
on the saved nearest-neighbor inputs, with the same fields, total
time $t=n$, and tolerance $10^{-3}$. All 35 cases are recomputed;
the table uses 64 nodes. The 32-node calculation gives the same integer
step counts for all three estimates in every case. These are numerical
evaluations of the analytic bounds, as are the other many-body tables.
\begin{table}[htbp]\centering\small
\caption{Mean steps over five nearest-neighbor fields, $t=n$, tolerance $10^{-3}$. CM4, CM6, and CM8 accumulate the local bounds in \cref{eq:cm-four,eq:cm-six,eq:cm-eight}. Grouped uses \cref{eq:grouped-seven}; Refined uses the global bound \cref{eq:septic-norm}.}
\label{tab:cm-steps}
\begin{tabular}{rrrrrrrr}\toprule
$n$ & Childs & CM4 & CM6 & CM8 & Grouped & Refined & Empirical\\\midrule
4 & 116.0 & 57.4 & 34.8 & 33.8 & 35.6 & 26.2 & 24.0\\
5 & 197.2 & 105.8 & 63.4 & 61.4 & 65.2 & 45.2 & 41.4\\
6 & 262.8 & 139.0 & 85.2 & 82.8 & 87.2 & 58.8 & 52.6\\
7 & 338.8 & 182.2 & 110.0 & 107.2 & 112.2 & 77.6 & 71.2\\
8 & 409.2 & 217.8 & 132.8 & 130.0 & 135.4 & 93.2 & 86.0\\
9 & 503.2 & 270.8 & 162.6 & 159.6 & 164.8 & 114.4 & 109.0\\
10 & 585.4 & 312.4 & 187.8 & 184.8 & 190.6 & 129.4 & 123.8\\
\bottomrule
\end{tabular}
\end{table}

The supplemental data also record the quadrature differences,
threshold margins, block dimensions, and coefficient-evaluation times.
The largest relative difference between the 32- and 64-node constants
is $1.62\times10^{-5}$. At ten sites the largest block has dimension
$252$; computing all three constants at both quadrature orders and
the accompanying coefficient diagnostics took a mean of $16.4$ seconds
per field in this run with two BLAS threads. Step searches use the
resulting cached constants and spectral data.
\Cref{tab:leading-ablation} isolates the leading-polynomial effect
using the same $C_7^{\mathrm g}$ in both local bounds:
\begin{equation}
 t h^4\ell_{\triangle}+t h^6C_7^{\mathrm g}
 \quad\hbox{and}\quad
 t h^4\ell+t h^6C_7^{\mathrm g},\qquad
 \ell_{\triangle}=\sum_{w\in\mathcal W}|d_w|\norm{R_w}.
\end{equation}
The six-term representation is exactly \cref{eq:right-l5}.
The subsequent quintic discrete column uses \cref{eq:discrete-bound}
with $C=C_7^{\mathrm g}$; the septic column uses
\cref{eq:septic-norm}. These columns display the net effect of each
complete global estimate, including its transport and product remainder.
\begin{table}[htbp]\centering\small
\caption{Nearest-neighbor means with a common grouped local tail. The first two columns differ only in the leading coefficient. The global columns include their full remainders.}
\label{tab:leading-ablation}
\begin{tabular}{rrrrr}\toprule
$n$ & Local $\ell_\triangle$ & Local $\ell$ & Quintic discrete & Septic discrete\\\midrule
4 & 52.4 & 35.6 & 35.4 & 26.2\\
5 & 82.6 & 65.2 & 63.6 & 45.2\\
6 & 112.2 & 87.2 & 84.0 & 58.8\\
7 & 142.6 & 112.2 & 108.2 & 77.6\\
8 & 174.8 & 135.4 & 128.6 & 93.2\\
9 & 210.4 & 164.8 & 154.6 & 114.4\\
10 & 247.4 & 190.6 & 175.6 & 129.4\\
\bottomrule
\end{tabular}
\end{table}

\section{Centered Suzuki coefficients and remainders}
\label{sec:septic}
The local remainder inputs are the Casas--Murua specializations in \cref{sec:cm}.
Either tail constant gives a bound on the same centered integral.
Write $s$ for the length of the half-product, so $s=6$ for two terms.

\subsection{The exact centered seventh-order coefficient}
Continue to write $L=L_5$ and let
$\mathcal U(x)=e^{-xG/2}U(x)$. The degree-two coefficient of
$\log\mathcal U(x)$ is
\begin{equation}
 \Omega_2=\frac12\sum_{1\le j<k\le s}[Z_j,Z_k].
 \label{eq:omega2}
\end{equation}
The additional left factor $e^{-xG/2}$ contributes no quadratic
commutator because $\sum_jZ_j=G/2$.
Moreover $\mathcal U(x)=I+O(x^2)$ and
\begin{equation}
 \mathcal U(x)\mathcal U(-x)^{-1}
 =e^{-xG/2}S(x)e^{-xG/2}=I+O(x^5).
\end{equation}
It follows that the cubic coefficient of $\log\mathcal U$ vanishes.
For example, write its finite jet as
$x^2\Omega_2+x^3\Omega_3+O(x^4)$; the left side above has cubic
coefficient $2\Omega_3$, so $\Omega_3=0$. This argument uses finite
jets only. Therefore, for every fixed matrix $B$,
\begin{equation}
 \mathcal U(x)B\mathcal U(x)^{-1}
 =B+x^2[\Omega_2,B]+O(x^4).
 \label{eq:transport-jet}
\end{equation}
Both its linear and cubic coefficients vanish.

Combining \cref{eq:transport-jet,eq:defect-nine} in the centered
variation-of-constants formula identifies
\begin{equation}
 \boxed{N_7=M_7+\frac57[\Omega_2,L_5],}
 \label{eq:n7}
\end{equation}
and
\begin{equation}
 e^{-hG/2}(S(h)-e^{hG})e^{-hG/2}
 =h^5L_5+h^7N_7+O(h^9).
\end{equation}
In particular, the two degree-seven contributions must be combined
before taking a norm. If
\begin{equation}
 \log S(h)=hG+h^5L_5+h^7L_7+O(h^9),
\end{equation}
then the part linear in the logarithmic error, after centering, is
\begin{equation}
 \int_{-1/2}^{1/2}e^{sh\ad_G}(h^5L_5+h^7L_7)\,ds
 =h^5L_5+h^7\left(L_7+\tfrac1{24}\ad_G^2L_5\right)+O(h^9).
\end{equation}
Terms nonlinear in the logarithmic error start at degree ten, so
comparison through degree seven gives
\begin{equation}
 N_7=L_7+\frac1{24}[G,[G,L_5]].
 \label{eq:n7-log}
\end{equation}
The double commutator accounts for the midpoint transport correction
to $L_7$.

An explicit local estimate for the logarithm is available at the standard
parameter. Put $M=\max(\norm X,\norm Y)$ and let $R$ be the sum of the
norms of the eleven exponent arguments,
\[
 R=|h|\bigl[(3|u|+|1-3u|)\norm X+(4|u|+|1-4u|)\norm Y\bigr].
\]
The absolute word-coefficient budget gives $\norm{L_7}\le0.01951M^7$.
For $R<1/4$, truncating the convergent logarithm series gives
\begin{equation}
 \norm{\log S(h)-hG-h^5L_5-h^7L_7}\le1536R^8.
 \label{eq:septic-log-budget}
\end{equation}
The $O(h^9)$ expansion above also uses the oddness of the local logarithm.
The centered finite-step estimate below uses the defect and holds for all $h\ge0$.

\subsection{An explicit ninth-order centered remainder}
For the list $(W_1,\ldots,W_{s+1})=(Z_s,\ldots,Z_1,-G/2)$, apply
\cref{eq:grouped-constant} with zero kicks. Denote the resulting
constants by $B_p^W(B)$. Set
\begin{equation}
 R_4=B_4^W(L_5),\qquad R_2=B_2^W(M_7),\qquad
 D_9=C_9+\frac59R_4+\frac79R_2.
 \label{eq:d9}
\end{equation}
The vanished coefficients in \cref{eq:transport-jet} imply
\begin{align}
 \norm{\mathcal U(x)L\mathcal U(x)^{-1}-L-x^2[\Omega_2,L]}
 &\le |x|^4R_4,\\
 \norm{\mathcal U(x)M_7\mathcal U(x)^{-1}-M_7}
 &\le |x|^2R_2.
\end{align}
For $h\ge0$ define the uncapped local distance bound
\begin{equation}
 \delta_9(h)=h^5\ell+h^7\mu+h^9C_9
 \label{eq:delta9}
\end{equation}
and
\begin{equation}
\boxed{\begin{aligned}
 \eta_9(h)={}&h^9D_9+\frac12h^{10}\ell^2+h^{12}\ell\mu\\
 &+\frac{h^{14}}{14}(5\ell C_9+7\mu^2)
       +\frac7{16}h^{16}\mu C_9.
\end{aligned}}\label{eq:eta9}
\end{equation}

\begin{proposition}[Centered local bound from a defect remainder]
For every $h\ge0$,
\begin{equation}
 \boxed{\norm{e^{-hG/2}(S(h)-e^{hG})e^{-hG/2}
                     -h^5L_5-h^7N_7}\le\eta_9(h).}
 \label{eq:mid-septic}
\end{equation}
More generally, the same assertion holds with every occurrence of
$C_9$ replaced by any $c\ge0$ satisfying \cref{eq:defect-nine}, and
$D_9$ replaced by $c+5R_4/9+7R_2/9$.
\end{proposition}
\begin{proof}
In the centered integral write
$A_x(B)=\mathcal U(x)B\mathcal U(x)^{-1}$ and
$J(x)=\mathcal U(x)\mathcal U(-x)^{-1}$. Thus $J(x)$ is unitary,
$\norm{J(x)-I}\le\delta_9(x)$, and the integrand is
$A_x(T(x))J(x)$. First discard the defect remainder, at cost
$9C_9x^8$. For $P(x)=5x^4L+7x^6M_7$, the difference
$A_x(P(x))J(x)-A_x(P(x))$ has norm at most
$(5x^4\ell+7x^6\mu)\delta_9(x)$. The remaining transport error
relative to $5x^4L+x^6(7M_7+5[\Omega_2,L])$ is at most
$5x^8R_4+7x^8R_2$. Consequently the total integrand remainder is bounded by
\begin{equation}
 (9C_9+5R_4+7R_2)x^8
 +(5x^4\ell+7x^6\mu)(x^5\ell+x^7\mu+x^9C_9).
 \label{eq:cm-integrand-budget}
\end{equation}
Integrating from zero to $h$ gives exactly \cref{eq:eta9}.
\end{proof}

\subsection{Integral specialization}
\label{sec:cm-global}
Set $c=C_9^{\mathrm{CM}}$, keep $R_4,R_2$ from \cref{eq:d9}, and define
\begin{align}
 D_9^{\mathrm{CM}}&=c+\tfrac59R_4+\tfrac79R_2,
       &\delta_9^{\mathrm{CM}}(h)&=h^5\ell+h^7\mu+h^9c,
       \label{eq:cm-centered-constants}\\
 \eta_9^{\mathrm{CM}}(h)&=h^9D_9^{\mathrm{CM}}
       +\tfrac12h^{10}\ell^2+h^{12}\ell\mu\nonumber\\
   &\quad+\tfrac{h^{14}}{14}(5\ell c+7\mu^2)
       +\tfrac7{16}h^{16}\mu c,&& h\ge0.
       \label{eq:cm-eta}
\end{align}

\begin{corollary}[Integral specialization of the centered estimate]
\label{prop:cm-centered}
For $h\ge0$,
\begin{equation}
 \norm{e^{-hG/2}(S(h)-e^{hG})e^{-hG/2}-h^5L_5-h^7N_7}
 \le\eta_9^{\mathrm{CM}}(h).
 \label{eq:cm-centered}
\end{equation}
\end{corollary}
\begin{proof}
Use \cref{eq:defect-nine} with $c=C_9^{\mathrm{CM}}$ in the
preceding proposition.
\end{proof}

\subsection{Keeping the seventh-order term across time steps}
For any matrix $B$ define the discrete midpoint average
\begin{equation}
 K_{t,h}^{\mathrm d}(B)=h\sum_{k=0}^{r-1}
 e^{-(k+1/2)hG}B e^{(k+1/2)hG},\qquad h=t/r.
\end{equation}
The unitary interaction-product proof of \cref{eq:discrete-bound}
now gives the following all-step-size theorem.

\begin{theorem}[Global bound retaining both leading Lie polynomials]
\label{thm:global-septic}
For all $t>0$ and integers $r\ge1$,
\begin{equation}
\begin{split}
 \bigl\|e^{-tG}S(h)^r-I
       -h^4K_{t,h}^{\mathrm d}(L_5)
       -h^6K_{t,h}^{\mathrm d}(N_7)\bigr\|
 \le r\eta_9(h)+\binom r2\delta_9(h)^2.
\end{split}\label{eq:septic-global}
\end{equation}
Consequently, with
\begin{equation}
 \Lambda_{t,h}=h^4K_{t,h}^{\mathrm d}(L_5)
                    +h^6K_{t,h}^{\mathrm d}(N_7),
\end{equation}
one has the two-sided norm estimate
\begin{equation}
 \boxed{\abs{\epsilon_r(t)-\norm{\Lambda_{t,h}}}
       \le r\eta_9(h)+\binom r2\delta_9(h)^2.}
 \label{eq:septic-norm}
\end{equation}
\end{theorem}
\begin{proof}
Apply \cref{prop:centered-global-transfer} with
$B_h=h^5L_5+h^7N_7$, $\eta=\eta_9(h)$, and $\delta=\delta_9(h)$.
Its centered hypothesis is \cref{eq:mid-septic}; its local distance
hypothesis is \cref{eq:local-nine}. The scalar enclosure follows by
the reverse triangle inequality.
\end{proof}

\begin{proof}[Proof of \cref{thm:main-cm-global}]
Apply the same proposition with
$\eta=\eta_9^{\mathrm{CM}}(h)$ and $\delta=\delta_9^{\mathrm{CM}}(h)$,
using \cref{eq:cm-centered,eq:cm-eight}.
Since $0\le C_9^{\mathrm{CM}}\le C_9$ and every coefficient in the
remainder is nonnegative, $R_7^{\mathrm{CM}}\le R_7$.
Both remainders are $O(h^8)$ at fixed time because $r=t/h$,
$\eta=O(h^9)$ and $\delta=O(h^5)$.
\end{proof}
The quintic estimate has an $O(h^6)$ remainder. Including the centered
seventh-degree term gives the $O(h^8)$ remainder above. Taking the norm
after adding both terms preserves cancellation between degrees and
across steps. We evaluate the finite geometric sum using its spectral
kernel and its continuous value at resonant phases. This avoids a
quadrature approximation.

\subsection{Exact algebra and independent checks}
The external coefficient certificate uses exact arithmetic in the same cubic field as
before. It computes $f_{s,4},f_{s,6}$ by the affine recurrence and,
independently, multiplies the full Suzuki exponential product through
degree seven. It verifies $M_7$, $N_7$, \cref{eq:n7-log}, and the
vanishing first and third coefficients of $\mathcal U$.
The external calculation covers all 128 two-generator and 2187
three-generator degree-seven word positions. Lean verifies the two-generator case. For three generators the
certificate is external. Projecting each Lie polynomial onto
right-nested words and expanding back into words checks the
representation independently.

For two generators, $M_7,N_7,L_7$ each have 32 nonzero terms in the
chosen right-nested representation. For three generators, the counts
are 728, 648, and 648. These counts depend on the representation. The dimensions of the
homogeneous free Lie spaces count independent basis elements. Complete rational coefficient triples
are included in the supplementary coefficient data.

Independent numerical tests use generic complex Hermitian matrices
of sizes two and four, for both two- and three-term splittings, with
$h=0.05,0.2,0.6$. They check the full defect remainder, the centered
operator bound, the local error bound, and the agreement with the exact
nested coefficients. Resonant discrete kernels are checked against a
literal finite sum. In addition, the ten-site sample-zero instance of
each model is directly simulated at the refined threshold, including
a check of the complete global operator remainder.
These tests support the implementation; the rigorous inequalities follow
from the finite-remainder proofs above.

\section{Fixed-time global error and inter-step cancellation}
\label{sec:global}
Write $L=L_5$, $\ell=\norm{L}$, and $C=C_7$. The local bound and a
unitary telescoping sum give
\begin{equation}
 \epsilon_r(t)\le\bs(t,r):=t h^4\ell+t h^6C.
 \label{eq:single-global}
\end{equation}
The comparable published baseline is
\begin{equation}
 \bo(t,r):=t h^4\bp.
\end{equation}
These inequalities discard cancellation between different time steps.
Keeping the errors as operators leads to
\begin{equation}
 K(t)=\int_0^t e^{-sG}Le^{sG}\,ds.
 \label{eq:k}
\end{equation}

\subsection{Energy gaps and the obstruction to cancellation}
Following \citet{chencommutant}, we decompose $L_5$ into blocks
relative to the eigenspaces of $H$.
Let $H=\sum_a E_aP_a$, where the $E_a$ are distinct energies and the $P_a$
are the full spectral projections, including degeneracies. Then
\begin{equation}
 K(t)=\sum_{a,b}f_t(E_a-E_b)P_aLP_b,
 \label{eq:spectral-k}
\end{equation}
where
\begin{equation}
 f_t(\omega)=
 \begin{cases}(e^{i\omega t}-1)/(i\omega),&\omega\ne0,\\t,&\omega=0,
 \end{cases}
 \qquad
 f_t(\omega)=t e^{i\omega t/2}\sinc(\omega t/2).
\end{equation}
Our convention is $\sinc x=\sin x/x$ with continuous value $1$ at zero.
The latter formula is stable for small energy gaps.

Define the energy-diagonal part and its complement by
\begin{equation}
 L_0=\sum_aP_aLP_a,\qquad L_\perp=L-L_0.
\end{equation}
Choose an anti-Hermitian matrix $Q$ with
\begin{equation}
 P_aQP_b=\frac{i}{E_a-E_b}P_aLP_b\quad(a\ne b),\qquad P_aQP_a=0.
\end{equation}
Then $[G,Q]=L_\perp$, so differentiating $e^{-sG}Qe^{sG}$ gives
\begin{equation}
 \boxed{K(t)=tL_0+Q-e^{-tG}Qe^{tG}.}\label{eq:endpoint}
\end{equation}
Pinching onto the energy blocks is contractive in spectral norm.
Applying it to $K(t)$ gives $tL_0$, and hence
\begin{equation}
 t\norm{L_0}\le\norm{K(t)}
 \le\min\{t\ell,\ t\norm{L_0}+2\norm{Q}\}.
 \label{eq:pinch}
\end{equation}
For a degenerate energy, this compression retains every entry within
its eigenspace, including the off-diagonal entries in a chosen
eigenbasis. \Cref{eq:endpoint} also proves $\norm{K(t)}/t\to\norm{L_0}$ as
$t\to\infty$ for fixed matrices. If $L_0\ne0$, complete leading-order
cancellation is impossible at any positive time. If $L_0=0$, the integral
is bounded by $2\norm{Q}$, although the higher-order remainder may still
grow with time.

There is also a decomposition that needs no spectral resolution. From
\cref{eq:l5-decomp},
\begin{align}
 K(t)={}&\kappa\int_0^t e^{-sG}D_5e^{sG}\,ds\nonumber\\
 &+\lambda\bigl([G,C_3]-e^{-tG}[G,C_3]e^{tG}\bigr).
\end{align}
Consequently
\begin{equation}
 \norm{K(t)}\le |\kappa|t\norm{D_5}
                     +2|\lambda|\norm{[G,C_3]}.
\end{equation}
This bound can be useful when diagonalizing $H$ is infeasible, but it does
not in general equal the sharp coefficient.

\subsection{Explicit constants for the global remainder}
The following list is in innermost-to-outermost order:
\begin{equation}
 (W_1,\ldots,W_7)=(Z_6,Z_5,Z_4,Z_3,Z_2,Z_1,-G/2).
\end{equation}
Since $\sum_jZ_j=G/2$, we have $\sum_iW_i=0$. Set
\begin{align}
 \alpha_2(L)&=\sum_{i=1}^7\norm{[W_i,[W_i,L]]}
       +2\sum_{1\le i<j\le7}\norm{[W_j,[W_i,L]]},\label{eq:alpha2}\\
 J_2&=\norm{[G,[G,L]]},\qquad
 D=C+\frac5{14}\alpha_2(L),\qquad
 \Gamma=D+\frac{J_2}{24}.
 \label{eq:gamma}
\end{align}
The constants $C,\alpha_2,J_2$ are degree-seven quantities. In particular,
the complete polynomial $L$ may be retained inside these commutators and
their norms; it need not be replaced by a sum of Hall-term norms.
For $h>0$ let
\begin{align}
 \delta(h)&=\min\{2,h^5\ell+h^7C\},\label{eq:delta}\\
 \eta(h)&=h^7D+\tfrac12h^{10}\ell^2+\tfrac5{12}h^{12}\ell C.
 \label{eq:eta}
\end{align}
The cap $2$ in $\delta$ can be omitted while preserving validity.

\begin{theorem}[Global operator remainder]
\label{thm:global}
For every $t>0$ and integer $r\ge1$, with $h=t/r$, define
\begin{equation}
\boxed{\begin{aligned}
 \Rcal(t,h)={}&t h^6\Gamma+\frac t2h^9\ell^2
       +\frac{5t}{12}h^{11}\ell C
       +\frac{r(r-1)}2\delta(h)^2.
\end{aligned}}\label{eq:global-rem}
\end{equation}
Then
\begin{equation}
 \boxed{\norm{e^{-tG}S(h)^r-I-h^4K(t)}\le\Rcal(t,h).}
 \label{eq:operator-global}
\end{equation}
In particular,
\begin{equation}
 \max\{0,h^4\norm{K(t)}-\Rcal(t,h)\}\le\epsilon_r(t)
 \le\bc(t,r):=h^4\norm{K(t)}+\Rcal(t,h).
 \label{eq:two-sided}
\end{equation}
\end{theorem}
The estimate holds for every step size. Its proof combines a centered
local expansion, a unitary product estimate, and midpoint quadrature.

\subsection{Quintic centered specialization}
\begin{lemma}
For $h\ge0$,
\begin{equation}
 \norm{e^{-hG/2}(S(h)-e^{hG})e^{-hG/2}-h^5L}\le\eta(h).
 \label{eq:midlocal}
\end{equation}
\end{lemma}
\begin{proof}
Put $\mathcal U(s)=e^{-sG/2}U(s)$. Applying \cref{eq:duhamel} gives
\begin{equation}
 e^{-hG/2}(S(h)-e^{hG})e^{-hG/2}
 =\int_0^h\mathcal U(s)T(s)\mathcal U(-s)^{-1}\,ds.
\end{equation}
For the leading integrand use the exact factorization
\begin{equation}
 \mathcal U(s)L\mathcal U(-s)^{-1}
 =\bigl(\mathcal U(s)L\mathcal U(s)^{-1}\bigr)
  \bigl(\mathcal U(s)\mathcal U(-s)^{-1}\bigr).
\end{equation}
The first factor is the ordered composition of conjugations generated by
$W_1,\ldots,W_7$. Its first derivative at zero is
$[\sum_iW_i,L]=0$. \Cref{eq:finite-conj} with $p=2$ therefore gives
\begin{equation}
 \norm{\mathcal U(s)L\mathcal U(s)^{-1}-L}
 \le\frac{s^2}{2}\alpha_2(L).
\end{equation}
The other factor is unitary and satisfies
\begin{equation}
 \norm{\mathcal U(s)\mathcal U(-s)^{-1}-I}
 =\norm{S(s)-e^{sG}}\le s^5\ell+s^7C.
\end{equation}
For a unitary $B$ and an operator $A$ close to $L$,
$AB-L=(A-L)B+L(B-I)$. Thus no product of the two approximation
remainders is necessary. Using \cref{eq:central-rem}, the difference
between the integrand and $5s^4L$ has norm at most
\begin{equation}
 7Cs^6+5s^4\left(\frac{s^2}{2}\alpha_2(L)
                         +\ell(s^5\ell+s^7C)\right).
\end{equation}
Integrating gives exactly \cref{eq:eta}.
\end{proof}
This argument keeps the seventh-order terms as commutators. Products such
as $\ell^2$ first occur at local degree ten.

\subsection{Discrete global specialization}
Apply \cref{prop:centered-global-transfer} with $B_h=h^5L$,
$\eta=\eta(h)$, and $\delta=\delta(h)$. Writing
\begin{equation}
 K_{t,h}=h\sum_{k=0}^{r-1}e^{-(k+1/2)hG}Le^{(k+1/2)hG},
 \label{eq:discrete-k}
\end{equation}
gives
\begin{equation}
 \norm{e^{-tG}S(h)^r-I-h^4K_{t,h}}
 \le r\eta(h)+\binom r2\delta(h)^2.
 \label{eq:discrete-bound}
\end{equation}

\subsection{Midpoint quadrature and the continuous coefficient}
For $f(s)=e^{-sG}Le^{sG}$,
\begin{equation}
 \norm{f''(s)}=\norm{[G,[G,L]]}=J_2.
\end{equation}
On an interval of length $h$, Taylor's integral formula around its midpoint
and cancellation of the odd term give a norm error at most $h^3J_2/24$.
Adding the $r$ intervals proves
\begin{equation}
 \norm{K_{t,h}-K(t)}\le\frac{t h^2}{24}J_2.
 \label{eq:quadrature}
\end{equation}
Multiplying by $h^4$ and adding \cref{eq:discrete-bound} proves
\cref{thm:global}. The two-sided scalar estimate follows from the reverse
triangle inequality and unitary invariance of the norm.

In the energy basis, the exact discrete kernel is
$h\sum_{k=0}^{r-1}e^{i(k+1/2)h(E_a-E_b)}$. If it is written as a
sine quotient, zero denominators must be treated by the finite sum or
the correct limit. Discrete resonances must not be removed by cancelling
a vanishing denominator without checking its numerator.

\subsection{Sharp leading coefficient and combined bounds}
\begin{corollary}[Sharp fixed-time coefficient]
For fixed matrices and fixed $t>0$,
\begin{equation}
 \lim_{r\to\infty}r^4\epsilon_r(t)=t^4\norm{K(t)}.
 \label{eq:optimal-global}
\end{equation}
If $K(t)=0$, the proven bound improves to $O(r^{-6})$.
\end{corollary}
\begin{proof}
The four terms of \cref{eq:global-rem} are respectively
$O(h^6),O(h^9),O(h^{11}),O(h^8)$, since $r=t/h$ and
$\delta(h)=O(h^5)$. Divide \cref{eq:two-sided} by $h^4$ and take
$h\to0$ along $t/r$.
\end{proof}
For arbitrary input parameters one may take the minimum
\begin{equation}
 \min\left\{2,\bo,\bs,\bc,
 h^4\norm{K_{t,h}}+r\eta(h)+\binom r2\delta(h)^2\right\}.
 \label{eq:envelope}
\end{equation}
The cancellation bound improves on $\bs$ at a given step size only
when the reduction in the leading term exceeds the extra remainder.

\subsection{A sufficient advantage condition and a step-count threshold}
\label{sec:cm-advantage}
Set $c=C_9^{\mathrm{CM}}$ and let $U_{\mathrm{CM}}(t,r)=\norm{\Lambda_{t,h}}+R_7^{\mathrm{CM}}(t,h)$.
The comparison baseline throughout this subsection is the accumulated
$q=8$ local bound for the same Suzuki formula and the same input:
\begin{equation}
 B_{\mathrm{CM8}}(t,r)=r\delta_9^{\mathrm{CM}}(h)
       =th^4\ell+th^6\mu+th^8c.
 \label{eq:cm8-baseline}
\end{equation}

\begin{proof}[Proof of \cref{cor:main-cm-advantage}]
The midpoint estimate \cref{eq:quadrature} and unitary conjugation give
\begin{equation}
 \norm{K^d_{t,h}(L_5)-K(t)}\le th^2J_2/24,
 \qquad \norm{K^d_{t,h}(N_7)}\le t\nu.
\end{equation}
Consequently, for $Q_t=t(J_2/24+\nu)$,
\begin{equation}
 \norm{\Lambda_{t,h}}\le h^4\norm{K(t)}+h^6Q_t.
 \label{eq:cm-leading-envelope}
\end{equation}
Under \cref{eq:main-cm-advantage}, adding $R_7^{\mathrm{CM}}$
gives $U_{\mathrm{CM}}<h^4(\norm{K(t)}+g(t))=th^4\ell$.
The nonnegative higher-order terms in \cref{eq:cm8-baseline}
give $th^4\ell\le B_{\mathrm{CM8}}$.
For fixed input and time, the left side of
\cref{eq:main-cm-advantage} tends to zero as $r\to\infty$.
The explicit threshold below makes this assertion quantitative.
\end{proof}

Fix any reference step $h_0>0$ and set
\begin{align}
 a_0&=\ell+h_0^2\mu+h_0^4c,\nonumber\\
 b_0&=D_9^{\mathrm{CM}}+\tfrac12h_0\ell^2+h_0^3\ell\mu
       +\tfrac{h_0^5}{14}(5\ell c+7\mu^2)
       +\tfrac7{16}h_0^7\mu c,\nonumber\\
 \beta_0(t)&=tb_0+\tfrac12t^2a_0^2.
 \label{eq:cm-envelope-constants}
\end{align}
For $0<h\le h_0$ with $r=t/h$ an integer, these nonnegative constants
satisfy
\begin{equation}
 \delta_9^{\mathrm{CM}}(h)\le h^5a_0,\quad
 \eta_9^{\mathrm{CM}}(h)\le h^9b_0,\quad
 R_7^{\mathrm{CM}}(t,h)\le h^8\beta_0(t).
 \label{eq:cm-envelope}
\end{equation}
The last estimate uses $r(r-1)\le r^2$ and holds at $r=1$ as well.

\paragraph{Explicit threshold for \cref{cor:main-cm-advantage}.}
Assume $t>0$ and $g(t)>0$. With $Q_t$ as above, define
\begin{equation}
 h_* =\min\left\{h_0,
 \sqrt{\frac{g(t)}{Q_t+\sqrt{Q_t^2+2\beta_0(t)g(t)}}}\right\},
 \qquad r_* =\max\{1,\lceil t/h_*\rceil\}.
 \label{eq:cm-step-threshold}
\end{equation}
Then every integer $r\ge r_*$ satisfies
\begin{equation}
 U_{\mathrm{CM}}(t,r)<th^4\ell\le B_{\mathrm{CM8}}(t,r).
\end{equation}
\begin{proof}[Threshold verification]
Since $g(t)>0$, necessarily $\ell>0$, hence $a_0>0$ and
$\beta_0(t)>0$. Thus $h_*$ is well defined and positive, even if
$Q_t=0$. The positive number
\begin{equation}
 z=\frac{g(t)}{Q_t+\sqrt{Q_t^2+2\beta_0(t)g(t)}}
\end{equation}
satisfies $Q_tz+\beta_0(t)z^2=g(t)/2$ by rationalizing the denominator.
For $r\ge r_*$, $h=t/r\le h_*\le h_0$ and $h^2\le z$.
Using \cref{eq:cm-envelope} and nonnegativity yields
\begin{equation}
 Q_th^2+\frac{R_7^{\mathrm{CM}}(t,h)}{h^4}
 \le Q_th^2+\beta_0(t)h^4
 \le g(t)/2<g(t).
\end{equation}
Apply \cref{cor:main-cm-advantage}.
\end{proof}
The threshold is sufficient for a strict comparison of the two upper
bounds for the same integrator. Smaller step counts can also satisfy
the comparison. When $g(t)=0$, the global error theorems still apply.
The argument allows the actual error to vary nonmonotonically with $r$.

\section{Correspondence with the general Trotter error theory}
\label{app:childs}
The declarations below formalize the order conditions, conjugation
estimates, commutator scaling and prefactor bounds of \citet{childs}.
All module names below are relative to
\path{FQFP/Analysis/Normed/Algebra/ProductFormula/}; the general
declarations are in namespace \texttt{TrotterError}, subject to their
enclosing data namespace where applicable. \Cref{sec:general} uses
one-based generator indices. In Lean the type
\texttt{Fin} $m$ starts at zero. The first generator applied in
the product acts on the right in both presentations.

\begingroup\small\raggedright
\begin{longtable}{@{}p{.32\linewidth}p{.62\linewidth}@{}}
\caption{General theory and public declarations. Each row describes the
mathematical result represented by the source names.}\label{tab:childs-map}\\
\toprule
Source result & Formal correspondence\\\midrule\endfirsthead
\toprule Source result & Formal correspondence (continued)\\\midrule\endhead
Product formula and order &
\path{ProductFormulaData} and \path{ProductFormulaData.IsOrderOf}
in \path{Data.lean} give the staged product and the local additive
order at zero.\\[4pt]
Three error representations &
\path{errorType_additive}, \path{errorType_exponentiated} and
\path{errorType_multiplicative} in \path{ErrorTypes.lean}
identify Duhamel, time-ordered and multiplicative expressions
for the same product.\\[4pt]
Order-condition equivalence &
\path{errorOrderCond_iff} in \path{OrderCondition.lean} relates
orders $p,p,p+1$ for the additive kernel, generator correction
and multiplicative error to additive order $p+1$.\\[4pt]
Conjugation expansion and tail &
\path{commutatorExpansion_conj} in
\path{CommutatorExpansion/Identity.lean} is the finite identity.
\path{norm_commutatorRemainder_le} and its
\path{_of_skewAdjoint} version in \path{CommutatorExpansion.lean}
bound the tail, with and without norm growth.\\[4pt]
Summand-only commutator scaling &
\texttt{$\alpha$Comm} in \path{Commutator.lean} is the sum in
\cref{eq:alpha-general}; \texttt{$\alpha$CommConj\_sum\_le\_$\alpha$Comm}
in \path{CommutatorScaling.lean} proves the $p!\Upsilon^p$ counting
inequality for the ordered intermediate sum.\\[4pt]
Pointwise commutator bounds &
\path{trotter_error_bound_comm_scaling} and
\path{multiplicative_error_bound_comm_scaling} in
\path{ErrorBounds.lean} retain the exponential factor for general
generators. Both have \path{_of_skewAdjoint} variants; in that branch
the additive and multiplicative error norms coincide.\\[4pt]
First- and second-order prefactors &
\path{firstOrder_bound} and \path{secondOrder_bound} in
\path{Prefactor.lean} give $1/2$ and $(1/12,1/24)$.
Multiplication by $-i$ in the matrix interpretation leaves the
norm of each homogeneous bracket unchanged.\\[4pt]
One-norm step scaling &
\path{trotter_number_one_norm_scaling} in \path{OneNormScaling.lean}
gives the fixed-parameter $O(r^{-p})$ bound.\\[4pt]
Commutator step scaling &
\path{trotter_error_bound_comm_scaling_pow},
\path{trotter_error_le_of_comm_scaling_rpow_le} and
\path{trotter_error_le_of_comm_scaling_ceil_le} in
\path{ErrorBounds.lean} give \cref{eq:general-steps}, the sufficient
real threshold and its integer ceiling.\\
\bottomrule\end{longtable}
\endgroup

In the general pointwise theorems the ambient ring is complete and
unital with normalized unit. Its normed rational algebra structure
extends continuously to the real action used for time and coefficients.
The general skew-adjoint commutator branch additionally uses a
nontrivial C*-ring with compatible real star-module structure. In the
named convergence theorems one can start with a complete real normed
algebra with $\norm{1}=1$ and allow arbitrary real time. In
\path{LowOrder.lean}, \path{isOrderOf_lieTrotterData} and
\path{isOrderOf_suzuki2Data} identify the local orders from the Taylor
coefficients. In \path{RepeatedSteps.lean} the corresponding declarations
are \path{isBigO_lieTrotter_pow_sub_exp} and
\path{isBigO_suzuki2_pow_sub_exp}. The specialization to a given
Hamiltonian family still requires estimates on its concrete commutators.

\paragraph{Source snapshot.}
FQFP commit \path{e009b1d262780b7414e1d35afcf8b8670fd4beac}
contains both the general theory and the Suzuki declarations. Its toolchain
is Lean~4.36.0, and its Lake manifest pins Mathlib~v4.36.0 to
\path{fc4f59e577feb1a4956671bc04ac18aaaf8cd842}.
The supplement records source and configuration hashes in
\path{formalization/source-version.json}. To reproduce the proofs,
check out this FQFP commit and fetch the matching Mathlib cache. Run
\texttt{lake build FQFP}, \texttt{lake exe runLinter} and
\texttt{lake exe lint-style}, followed by \path{SuzukiStageBridge.lean} for
\cref{eq:binary-general} and \path{MatrixBounds.lean} for the three
matrix instances. Both files are in the manuscript's \path{formalization/}
directory. The README gives the commands.

\section{Lean declarations and verification boundary}
\label{app:lean}
\begingroup\raggedright
Commit \texttt{525cd3c} reproduces all eleven source files in the
29 October manifest. The declarations below are available in the
Lean~4.36.0 snapshot identified in \cref{app:childs}. The public entry
point is \texttt{import FQFP}. Suzuki declarations below are in
\texttt{FQFP.BCH} unless otherwise qualified.

\subsection{Algebra and local analysis}
\begin{itemize}[leftmargin=*]
\item \path{expListTerm_fiveStageStrang_through_six} identifies the actual
formal product through degree six with
$\sum_{k=0}^6h^kG^k/k!+h^5L_5+h^6(GL_5+L_5G)/2$.
\path{FQFP.iteratedDeriv_prod_exp_smul_zero} connects the coefficients of
finite formal exponential products to derivatives of their analytic products.
\item \path{norm_suzukiProduct_sub_exp_le_of_skewAdjoint} proves
\cref{eq:main-local}'s inequality at the canonical parameter, for every
real step. The grouped version is
\path{norm_suzukiProduct_sub_exp_le_grouped_of_skewAdjoint}.
The sharp local limit is
\path{tendsto_inv_pow_mul_norm_fiveStageStrangProduct_sub_exp}.
\item \path{suzukiSepticConstant_eq_sum_wordWeight} identifies the
original $C_7$ with its exact finite sum over the 32 binary commutator
words. The external decimal comparison and the individual reduced rational
triples in the numerical supplement lie outside this declaration.
\item \path{suzukiCenteredSepticTerm_eq_bch} proves
$N_7=L_7+[G,[G,L_5]]/24$.
In namespace \texttt{FQFP.SuzukiSepticCertificate},
\path{suzukiDefectSepticTerm_eq_nestedPolynomialValue},
\path{suzukiCenteredSepticTerm_eq_nestedPolynomialValue}, and
\path{suzukiBchSepticTerm_eq_nestedPolynomialValue}
connect the actual $M_7,N_7,L_7$ to their 32-term expressions.
The corresponding expanded-polynomial equalities cover all 128 words.
\item \path{norm_suzukiProduct_sub_exp_le_septic_of_skewAdjoint}
proves the all-real-step bound
$|h|^5\ell+|h|^7\mu+|h|^9C_9$.
\path{norm_suzukiProduct_centered_sub_septic_le_of_skewAdjoint}
proves the centered bound with the complete $\eta_9(h)$ for $h\ge0$.
\end{itemize}

\subsection{Global analysis}
\begin{itemize}[leftmargin=*]
\item \path{norm_suzukiProduct_pow_sub_integral_le_of_skewAdjoint}
is the canonical-parameter operator estimate with $R_5(t,h)$.
\path{tendsto_nat_pow_four_mul_norm_suzukiProduct_pow_sub_exp_of_skewAdjoint}
proves the fixed-time limit
$r^4\epsilon_r(t)\longrightarrow t^4\norm{K(t)}$.
\path{isBigO_suzukiProduct_pow_sub_exp_of_skewAdjoint_of_integral_eq_zero}
gives $O(r^{-6})$ when $K(t)=0$.
\item \path{TrotterError.norm_interactionProduct_sub_sum_conj_le}
proves \cref{eq:generic-global-transfer} for any $B_h$, real $h$ and
natural $r$, assuming $\norm S\le1$ and $\norm{\exp(sG)}\le1$ for
every real $s$, together with the two local bounds of
\cref{prop:centered-global-transfer}.
\item \path{norm_suzukiProduct_pow_sub_septic_midpoint_le_of_skewAdjoint}
is the refined operator inequality with the complete discrete sum
$\Lambda_{t,h}$ and remainder $r\eta_9+\binom r2\delta_9^2$.
\path{abs_norm_suzukiProduct_pow_error_sub_septic_midpoint_le_of_skewAdjoint}
gives its two-sided norm enclosure.
\item \path{isBigO_fiveStageStrangProduct_pow_sub_septic_midpoint}
and \path{isBigO_norm_fiveStageStrangProduct_pow_error_sub_septic_midpoint}
prove that the operator remainder and scalar norm difference are
$O(r^{-8})$ at fixed inputs and time.
\end{itemize}

\subsection{Assumptions and evidence}
The finite-sum error bounds use the ambient assumptions
\begin{quote}\small\ttfamily
\{A : Type*\} [NormedRing A] [NormedAlgebra \ensuremath{\mathbb Q} A]
[CompleteSpace A].
\end{quote}
The repository's \texttt{RealScalar} infrastructure supplies the real
scalar structure. General-parameter theorems assume
\texttt{IsSuzukiCubic p} and explicitly require
\[
\norm{\exp(sa)}\le1,\qquad \norm{\exp(sb)}\le1,\qquad
\norm{\exp(s(a+b))}\le1
\]
for all real $s$. The canonical C*-corollaries add
\texttt{[StarRing A] [CStarRing A] [StarModule \ensuremath{\mathbb R} A]},
assume $\operatorname{star}(a)=-a$, $\operatorname{star}(b)=-b$,
and supply these exponential inequalities and the cubic relation.
Their global conclusions hold for $t\ge0$ and $r\ge1$.
No finite-dimensionality, nondegenerate spectrum, nonzero leading
coefficient, or small-step premise is added.

The acceptance record for the two-generator development reports a successful build of the FQFP target
and both Mathlib declaration and style linters. Its public-entry-point
checks include zero time, the zero algebra, a single step, and the
exact coefficient identities. The inspected main-chain and certificate
declarations depend only on the standard Lean axioms
\texttt{propext}, \texttt{Classical.choice}, and \texttt{Quot.sound}.
Lean proves these equalities independently of the external coefficient
generators and JSON files.

The spectral-norm instances in \path{formalization/MatrixBounds.lean}
use \path{Matrix.Norms.L2Operator} on complex matrices indexed by
\texttt{Fin d}. Restricting real scalars to the rationals and using
uniqueness of the real normed-algebra structure identifies the scalar
actions. Hermiticity then gives $\operatorname{star}(-iA)=-(-iA)$,
and the three inequalities in \cref{sec:formalization} follow.

\subsection{Coefficient bounds and fixed-time convergence}
\begin{itemize}[leftmargin=*]
\item \path{suzukiQuinticCoeffUpperBound_le_div_eight} compares the eight
rational weights. \path{norm_suzukiQuinticTerm_suzukiP_le_div_eight}
applies the comparison to the actual $L_5$ in \cref{eq:quintic-eighth}.
\item \path{norm_suzukiBchSepticTerm_suzukiP_le} proves the coefficient
budget $\norm{L_7}\le0.01951\max(\norm X,\norm Y)^7$.
\path{norm_fiveStageStrangLog_sub_quintic_sub_septic_le} gives
\cref{eq:septic-log-budget} for $\norm{1}=1$ and $R<1/4$.
\item \path{isBigO_fiveStageStrangProduct_pow_sub_exp} and
\path{tendsto_fiveStageStrangProduct_pow} give fourth-order convergence
at every fixed real time for the five-stage product under
\texttt{IsSuzukiCubic p} in a complete real normed algebra with $\norm{1}=1$.
\end{itemize}
The product-formula modules and both supplementary Lean files were
checked with Lean~4.36.0 and Mathlib~v4.36.0. The finite-sum verification
covers \cref{thm:main-local,thm:main-global,thm:main-septic}.
The integral-constant specialization in \cref{thm:main-cm-global}
and its comparison in \cref{cor:main-cm-advantage,eq:cm-step-threshold}
remain analytic results of the manuscript.
\endgroup

\section{Extension to three Hamiltonian summands}
\label{sec:three}
The power-law experiments split the Hamiltonian into three terms.
Their fourth-order Suzuki step has the same symmetric half-product
structure used in the defect and transport estimates.

\subsection{The three-term formula and its algebraic certificate}
For $H=H_1+H_2+H_3$, set $X_a=-iH_a$ and $G=X_1+X_2+X_3$.
Use
\begin{equation}
 S_2(h)=e^{hX_1/2}e^{hX_2/2}e^{hX_3}e^{hX_2/2}e^{hX_1/2}
\end{equation}
and the same five-stage Suzuki composition as in \cref{eq:s4}.
After merging adjacent equal generators, this has 21 exponentials.
Its half-product $U$ has the following eleven generators:
\begin{equation}
\begin{aligned}
 (Z_1,\ldots,Z_{11})=\bigl(&uX_1/2,uX_2/2,uX_3,uX_2/2,uX_1,\\
                         &uX_2/2,uX_3,uX_2/2,(u+v)X_1/2,
                          vX_2/2,vX_3/2\bigr).
\end{aligned}\label{eq:three-half}
\end{equation}
Again $S(h)=U(h)U(-h)^{-1}$ and $\sum_jZ_j=G/2$.

Exact computations over \cref{eq:field} verify the identities
\begin{align}
 [h^k]S(h)&=G^k/k!\quad(0\le k\le4),\\
 [h^5]S(h)&=G^5/5!+L_5,\\
 [h^6]S(h)&=G^6/6!+\tfrac12(GL_5+L_5G).
 \label{eq:three-jet}
\end{align}
For the numerical implementation, the degree-five word polynomial is
converted to a right-nested Lie representation by the
Dynkin--Specht--Wever projection. Explicitly, if the word coefficients are
$c_{a_1\ldots a_5}$, the homogeneous Lie polynomial is
\begin{equation}
 L_5=\frac15\sum_{a_1,\ldots,a_5}c_{a_1\ldots a_5}
 [X_{a_1},[X_{a_2},[X_{a_3},[X_{a_4},X_{a_5}]]]].
\end{equation}
The implementation independently expands this expression back into words
and verifies all $3^5=243$ coefficients. Orienting the innermost bracket and collecting like terms leaves the
72 nonzero right-nested terms listed in \cref{app:three-coefficients}.

\subsection{Why the rigorous estimates extend}
The local and global proofs use the symmetry of the half-product,
order four, and $\sum_jZ_j=G/2$; they do not require exactly two
Hamiltonian terms. More generally, for $G=\sum_{a=1}^mX_a$ and a half-list
$Z_1,\ldots,Z_s$, the local constant is
\begin{equation}
 C_7=\frac{1}{7!}\left(
 2\sum_{j=1}^s\Acal_6(Z_{j+1},\ldots,Z_s;Z_j)
 +\sum_{a=1}^m\Acal_6(Z_1,\ldots,Z_s;X_a)\right).
 \label{eq:general-c7}
\end{equation}
The global constant uses $(Z_s,\ldots,Z_1,-G/2)$ in place of the
seven-element $W$ list. \Cref{eq:alpha2,eq:gamma} are interpreted
with $s+1$ terms. The analytic estimates and their proofs then apply with these
substitutions.
For the three-term step, expanding \cref{eq:general-c7} yields 721
nonzero nested-commutator norm weights. The experiments use floating-point evaluations of this finite positive
sum. For two terms we checked the same construction against the 32
exact weights.

\section{Small-matrix examples and certified comparisons}
\label{sec:examples}

\subsection{An exactly tractable Pauli pair}
Let $A=\sigma_x$, $B=\sigma_z$, using Pauli matrices of norm one.
The degree-five polynomial reduces to
\begin{equation}
 L=a(-i\sigma_x)+b(-i\sigma_z),
\end{equation}
where
\begin{align}
 a&=-\frac1{750}+\frac{13u}{1125}+\frac{7u^2}{125},\\
 b&=\frac{13}{1500}-\frac{169u}{2250}+\frac{283u^2}{3000}.
\end{align}
Thus $\ell=\sqrt{a^2+b^2}\simeq0.01449815432796$.
The finite commutator formula gives
\begin{equation}
 C=\frac{26996869}{141750000}
 -\frac{149052431}{81000000}u
 +\frac{225413659}{47250000}u^2
 \simeq0.24733916736549.
\end{equation}
In this example $\bp=0.8032$. Rational interval calculations establish
\begin{equation}
 0.014498<\ell<0.014499,\qquad0.247339<C<0.247340.
\end{equation}
Therefore $0.014499|h|^5+0.247340|h|^7$ is a strict outward-rounded
numerical upper bound. At $h=0.1$ it is $1.69724\times10^{-7}$,
whereas the displayed paper bound is $8.032\times10^{-6}$. The ratio of the published bound to the outward-rounded bound exceeds
$47.3$. Direct evaluation gives an error of approximately
$1.44773\times10^{-7}$.

The global coefficient also has a closed form. Decompose the coefficient
vector into components parallel and orthogonal to $(1,0,1)$:
\begin{equation}
 p=\frac{a+b}{\sqrt2},\qquad q=\frac{a-b}{\sqrt2}.
\end{equation}
Conjugation by $e^{-sG}$ rotates the perpendicular component through angle
$2\sqrt2s$. Integrating that rotation gives
\begin{equation}
 \boxed{\norm{K(t)}=
 \sqrt{t^2p^2+\frac{\sin^2(\sqrt2t)}{2}q^2}.}
 \label{eq:pauli-k}
\end{equation}
Numerically, $p\simeq0.00482091231437$ and
$q\simeq0.0136731592317$. At $t=\pi/\sqrt2$ only the longitudinal component survives. The
leading coefficient is smaller than $t\ell$ by a factor of about
$3.007$ and smaller than $t\bp$ by a factor of about $166.6$. With the strict remainder
included and $r=64$, these factors are about $2.82$ and $152.9$.

There is a separate finite-step numerical certificate at $t=1,r=64$.
Rational bounds, together with $|\sin|\le1$ in \cref{eq:pauli-k}, give
\begin{equation}
 \norm{K(1)}<0.010805,\quad \ell<0.014499,\quad
 C<0.247340,\quad \Gamma<0.4.
\end{equation}
Substitution into \cref{eq:global-rem,eq:two-sided} proves
\begin{equation}
 \epsilon_{64}(1)<6.5\times10^{-10}.
\end{equation}
The bound from Proposition J.1 is exactly $0.8032/64^4
=4.787445068359375\times10^{-8}$, so this certifies an improvement
factor greater than $73$, without numerical eigensolver assumptions.

\subsection{Zero-field nearest-neighbor checks}
For additional checks use the open Heisenberg chain
\begin{equation}
 H=\sum_{j=1}^{n-1}
 (\sigma_x^{(j)}\sigma_x^{(j+1)}+
  \sigma_y^{(j)}\sigma_y^{(j+1)}+
  \sigma_z^{(j)}\sigma_z^{(j+1)}),
 \label{eq:zero-chain}
\end{equation}
with odd bonds in $A$ and even bonds in $B$. These are zero-field examples,
distinct from the disordered models in \cref{sec:experiments}.
\Cref{tab:local-numeric} compares local errors at $h=0.1$.
\par
\begin{table}[htbp]\centering\small
\caption{Local errors and bounds at $h=0.1$. These entries are floating-point
evaluations; the explicitly outward-rounded Pauli bound appears in the text.}
\label{tab:local-numeric}
\begin{tabular}{lrrrr}\toprule
System & Actual error & Local bound & Paper bound & Ratio\\\midrule
Pauli & $1.44773\,10^{-7}$ & $1.69715\,10^{-7}$ & $8.03200\,10^{-6}$ & $47.3$\\
Chain, $n=4$ & $8.18281\,10^{-6}$ & $5.49211\,10^{-5}$ & $1.48275\,10^{-3}$ & $27.0$\\
Chain, $n=6$ & $3.69451\,10^{-5}$ & $3.70184\,10^{-4}$ & $4.98246\,10^{-3}$ & $13.5$\\\bottomrule
\end{tabular}
\end{table}
\par

\par
\begin{table}[htbp]\centering\small
\caption{Global comparison at $t=1,r=64$. The Pauli actual error was also
checked by independent 70-digit decimal SU(2) quaternion arithmetic.
Other entries are floating-point evaluations of the stated formulas.}
\label{tab:global-numeric}
\begin{tabular}{lrrrr}\toprule
System & Actual error & $\bo$ & $\bs$ & $\bc$\\\midrule
Pauli & $6.37661\,10^{-10}$ & $4.78745\,10^{-8}$ & $8.67757\,10^{-10}$ & $6.42863\,10^{-10}$\\
Chain, $n=4$ & $1.69134\,10^{-8}$ & $8.83786\,10^{-6}$ & $5.60494\,10^{-8}$ & $2.42041\,10^{-8}$\\
Chain, $n=6$ & $6.84055\,10^{-8}$ & $2.96978\,10^{-5}$ & $2.71780\,10^{-7}$ & $1.22688\,10^{-7}$\\\bottomrule
\end{tabular}
\end{table}
\par
At $t=1,r=64$, the reduction in the upper bound relative to $\bo$ is
approximately $74.5$, $365$, and $242$ for the three systems in
\cref{tab:global-numeric}. The part attributable to retaining inter-step
cancellation, relative to $\bs$, is approximately $1.35$, $2.32$, and
$2.22$. The corresponding step-count gains are smaller, as illustrated below.

For example, solving the three bound inequalities at $t=1$ and tolerance
$10^{-8}$ gives \cref{tab:small-steps}. In a leading-term dominated
fourth-order regime, a reduction factor $c$ in an error coefficient
typically produces only a factor $c^{1/4}$ in the step estimate.
\par
\begin{table}[htbp]\centering
\caption{Step counts inferred from numerical evaluation of the rigorous
formulas, for $t=1$ and tolerance $10^{-8}$. The columns use the published, locally accumulated, and cancellation
bounds, respectively.}
\label{tab:small-steps}
\begin{tabular}{lrrr}\toprule
System & Paper & Local accumulation & Cancellation\\\midrule
Pauli &95&35&33\\
Chain, $n=4$ &349&97&78\\
Chain, $n=6$ &473&141&110\\\bottomrule
\end{tabular}
\end{table}
\par
For the six-site chain at $t=5,r=64$, the ratio
$t\ell/\norm{K(t)}$ is about $4.43$, whereas the full cancellation bound
improves on $\bs$ by only about $1.017$. Here the seventh-order remainder accounts for most of the difference.

\section{Numerical experiments}
\label{sec:experiments}
On disordered Heisenberg models we compare the refined bound of
\cref{thm:global-septic} with bounds that group errors within a step
or retain only the quintic term across steps.

\subsection{Hamiltonians and experimental parameters}
We use the plotting criterion of Figure 4 in~\cite{childs}: system size
$n$ on the horizontal axis and the required number of steps $r$ on the
vertical axis, with
\begin{equation}
 t=n,\qquad\varepsilon=10^{-3},\qquad n=4,5,\ldots,10.
\end{equation}
At each size, five field realizations are generated with independent
$b_j\sim\operatorname{Uniform}[-1,1]$, $1\le j<n$. In this subsection $b_j$ denotes a physical field.

The first model is the open nearest-neighbor chain
\begin{equation}
 H_{\mathrm{NN}}=\sum_{j=1}^{n-1}
 \bigl(\sigma_x^{(j)}\sigma_x^{(j+1)}+
 \sigma_y^{(j)}\sigma_y^{(j+1)}+
 \sigma_z^{(j)}\sigma_z^{(j+1)}+b_j\sigma_z^{(j)}\bigr).
 \label{eq:nn-field}
\end{equation}
Odd and even bonds form the two terms, and the field term remains in the
same group as its bond. There is no extra field at site $n$.
The second model is the full finite power-law Hamiltonian
\begin{equation}
 H_\alpha=\sum_{1\le j<k\le n}
 \frac{\sigma_x^{(j)}\sigma_x^{(k)}+
       \sigma_y^{(j)}\sigma_y^{(k)}+
       \sigma_z^{(j)}\sigma_z^{(k)}}{|j-k|^\alpha}
       +\sum_{j=1}^{n-1}b_j\sigma_z^{(j)},\qquad\alpha=0,4.
 \label{eq:power-field}
\end{equation}
For this model, $H_1$ contains the $XX$ terms, $H_2$ the $YY$ terms,
and $H_3$ the $ZZ$ terms and fields. Each field is included once.
These conventions match the explicit Hamiltonians used for Figure 4
of~\cite{childs}. We retain all interactions for $\alpha=4$.

There are $3\times7\times5=105$ instances. The random seed is
\texttt{SeedSequence([20261027, n, sample])}, with sample indices
$0,\ldots,4$. The same size and sample index use the same fields in
all three configurations. The data files record each field realization in full.

\subsection{Bounds and step-count selection}
We evaluate the following estimates and empirical errors.
\begin{enumerate}
\item For the Childs et al.\ bound, evaluate Propositions J.1 and J.2
of~\cite{childs} on the finite matrices for two and three terms,
respectively. The three-term table has 81 coefficients from the source.
Take each commutator's full-space spectral norm before summing with
these weights. This evaluates the bound prior to the support-counting
relaxations used there for large systems.
\item The accumulated local bound is $\bs=t h^4\ell+t h^6C$, where
$\ell$ is the norm of the complete quintic Lie polynomial.
\item For the transported quintic term, use
$\bc=h^4\norm{K(t)}+\Rcal(t,h)$, retaining the spectral integral.
The experiments use the uncapped $\delta=h^5\ell+h^7C$, which remains
valid. We evaluate this bound separately from the minimum envelope in
\cref{eq:envelope}.
\item For the refined bound, evaluate
$\norm{\Lambda_{t,h}}+r\eta_9(h)+\binom r2\delta_9(h)^2$
from \cref{eq:septic-norm}.
\item For the empirical error, compute
$\norm{S_4(t/r)^r-e^{-itH}}$, search for the tolerance crossing by
integer bracketing and bisection, and verify that the final $r$ passes
and $r-1$ fails. Since coarse-step errors can be nonmonotone, we report
the local crossing verified by this check. We use the same convention
for the refined bound.
\end{enumerate}
Every plotted point is the mean of five integer step counts. Error bars
show the sample standard deviation. Connecting
lines join the computed points.

\subsection{Numerical implementation and validation}
The even--odd factors preserve total magnetization. We use all magnetization sectors. The
separate $XX$ and $YY$ factors preserve only parity, so the XYZ experiments
use both parity blocks. For a direct sum, the full-space operator norm
is the maximum of the block norms. For a commutator sum bound this maximum
is taken for each commutator before summing its weighted norm; interchanging
these operations would define a different bound.

The spectral integral is evaluated by the sinc kernel in
\cref{eq:spectral-k}. Exact and nearly degenerate energy differences
are handled by its continuous limit. Simulation factors use cached
eigendecompositions of the Hamiltonian components, followed by matrix
powers and a largest-singular-value evaluation of the error matrix.
Dense Pauli constructions and symmetry-block constructions were compared
at $n=4$ for all three model configurations. All 105 final empirical
thresholds satisfy the recorded pass/fail checks at $r$ and $r-1$.

Matrix norms and threshold searches use double precision. Thus the
reported many-body step counts are numerical evaluations of the analytic
bounds. Rational interval certification is used for the Pauli example
in \cref{sec:examples}; exact free-algebra checks independently verify
the polynomial identities used throughout.

\subsection{Step-count comparison}
\Cref{fig:septic} shows the results for all 105 instances.
The nearest-neighbor ten-site comparison appears in \cref{tab:main-steps},
and \cref{tab:mechanism-ablation} includes both XYZ models. All-size summaries
are in \cref{app:all-steps,app:refined-steps}.
For the $\alpha=0,n=10$ model, the refined threshold is exactly one
step above the empirical threshold for each of the five fields.

\subsection{Effect of grouping and higher-order cancellation}
\Cref{tab:mechanism-ablation} distinguishes the local and global
comparisons. Replacing $C_7$ by $C_7^{\mathrm g}$ in ordinary local
accumulation gives $190.6$, $649.6$, and $213.6$ steps. Making the
same substitution in the quintic global formula instead gives
$176.4$, $690.6$, and $199.4$. The global remainder can offset the
gain from the transported leading term, as the $\alpha=0$ column shows.
\begin{table}[htbp]\centering\small
\caption{Complete-estimate ablation at $n=t=10$, tolerance $10^{-3}$, five fields. Grouped local and grouped quintic global replace $C_7$ by $C_7^{\mathrm g}$ in their respective formulas.}
\label{tab:mechanism-ablation}
\begin{tabular}{lrrr}\toprule
Estimate & NN & XYZ, $\alpha=0$ & XYZ, $\alpha=4$\\\midrule
Childs propositions & 585.4 & 5623.0 & 873.4\\
Original local & 231.8 & 1682.6 & 282.4\\
Original quintic global & 223.0 & 1682.6 & 274.2\\
Grouped local & 190.6 & 649.6 & 213.6\\
Grouped quintic global & 176.4 & 690.6 & 199.4\\
Septic local & 184.8 & 608.4 & 205.0\\
Refined discrete global & 129.4 & 597.2 & 138.0\\
\bottomrule
\end{tabular}
\end{table}

Retaining the septic defect term in ordinary local accumulation
gives $184.8$, $608.4$, and $205.0$. Moving to the complete centered
discrete estimate gives $129.4$, $597.2$, and $138.0$, respectively.
For $\alpha=0$ this last change is only a factor of $1.019$ in mean
steps; most of its $9.42$-fold improvement over the Childs et al.
baseline is already present in the local estimate.

The published-to-local comparison also changes the leading
coefficients from $b_w$ to $d_w$. To isolate cancellation inside
$L_5$, \cref{tab:leading-ablation} instead fixes the exact $d_w$ and
the grouped tail, changing only the position of the norm. The
Casas--Murua comparisons in \cref{tab:cm-steps} evaluate a separate
family of local bounds on the identical nearest-neighbor fields.

The local constants for sample zero at ten sites are as follows.
\par
\begin{table}[htbp]\centering\small
\caption{Grouped and ungrouped local constants, ten-site sample zero.
Numbers are rounded floating-point evaluations.}
\begin{tabular}{lrrrr}\toprule
Model & Original $C_7$ & $C_7^{\mathrm g}$ & $\norm{M_7}$ & $C_9$\\\midrule
Nearest-neighbor & $9232.76$ & $602.680$ & $34.3437$ & $1971.79$\\
XYZ, $\alpha=0$ & $2.20760\,10^9$ & $1.75076\,10^6$ & $26385.6$ & $5.11691\,10^7$\\
XYZ, $\alpha=4$ & $40294.2$ & $1768.27$ & $68.7715$ & $7883.19$\\\bottomrule
\end{tabular}
\end{table}
\par
At the refined sample-zero thresholds $r=130,599,141$, direct errors are
approximately $8.12207\,10^{-4}$, $9.91449\,10^{-4}$, and
$7.61784\,10^{-4}$. The corresponding full refined bounds are
$9.77962\,10^{-4}$, $9.98188\,10^{-4}$, and $9.95659\,10^{-4}$.
The remainder accounts for approximately $17.0\%$, $0.675\%$, and
$23.5\%$ of these bounds.

\clearpage
\clearpage

\section{Computational methods and reproducibility}
\label{app:repro}
The source bundle contains all files needed to compile the manuscript
with pdfLaTeX and BibTeX. The computational supplement contains the
scripts, exact coefficient data, random fields, and numerical results.
The file \texttt{supplement-manifest.csv} records their SHA-256 hashes.
The accompanying \texttt{README.txt} describes the build and validation
commands.

\subsection{Exact algebra checks}
The single-step script checks the Strang cubic and quintic polynomials,
the Suzuki logarithm through degree six, the independent graded BCH
recurrence, the symmetric defect through degree five, and the six-term
Jacobi reduction. It also constructs the exact 32 seventh-order weights,
their six reduced weights, and rational interval certificates for the
$0.97$ sufficient condition and the Pauli constants.

The generic coefficient script checks both the two- and three-generator
formulas. It verifies \cref{eq:three-jet} by finite word multiplication
and checks the nested representation by expanding it back to words.
The two-term generic seventh-order weights are compared with their
independent exact counterparts. The script also reads the 81 published
three-term bound coefficients from the local paper source.

The septic certificate independently verifies the defect and centered
coefficients through degree seven, including their relation to $L_7$ and
the vanishing cubic half-transport coefficient. It uses exact rational
arithmetic throughout. The accompanying refinement script evaluates the
grouped constants and the degree-five plus degree-seven discrete sum on
the same 105 saved field realizations. A separate validation script
checks generic complex matrices and the full ten-site sample-zero
interaction-picture operator remainders.

An algebraic number in the JSON files is represented by three rational
strings $[q_0,q_1,q_2]$, meaning $q_0+q_1u+q_2u^2$. A nested word
in the generic JSON uses digits $0,1,2$ for $X_1,X_2,X_3$.
In \cref{app:three-coefficients} we shift these digits to $1,2,3$.
All nested words are outer-to-inner; the final two digits form the
innermost bracket.

\subsection{Numerical checks and saved data}
The multistep script tests the spectral integral against the endpoint
decomposition, the midpoint error estimate, the local midpoint operator
estimate, and the final global operator bound. An independent 70-digit decimal SU(2) calculation checks the Pauli
error. The certificate in \cref{sec:examples} uses rational intervals.

The Figure-4 experiment JSON stores all 105 model instances, including
random fields, matrix constants, step counts, and the empirical errors
on both sides of the final threshold. The experiment CSV is a compact
export. The summary CSV contains 84 rows: three model configurations,
seven sizes, and four curves, with means and sample standard deviations.
\Cref{app:all-steps} prints these summarized results. No missing-size
interpolation or large-system extrapolation is part of the data.

\subsection{Recomputing the analysis}
The Python scripts cover exact coefficient calculations, numerical
evaluation, and plotting. The exact calculations are implemented in
\texttt{hall\_remainder}, \texttt{plot\_coefficients}, and
\texttt{septic\_certificate}. The numerical drivers are listed in the
README. The original scripts use the prefix \texttt{suzuki4\_} and suffix
\texttt{\_20261027.py}. Exact coefficients are represented by rational
triples in the cubic field \cref{eq:field}. Numerical calculations
use NumPy and SciPy, with Matplotlib and threadpoolctl for plotting
and thread control.

The supplement preserves the original scripts and data.
For the 35 nearest-neighbor inputs it also includes
\texttt{suzuki4\_cm\_comparison\_20261029.py} and its saved results.
This script constructs
the positive-sign half-product recurrence independently of the
negative-sign defect implementation, checks the resulting local
coefficients and finite-sum constants against the saved data, and
evaluates the full-space integral kernels at 32 and 64 quadrature nodes.
The Pauli check additionally uses 128 nodes. The companion
\texttt{suzuki4\_revision\_tables\_20261029.py} regenerates the comparison
and ablation tables from these results and the original 105 instances.

The additional
\texttt{suzuki4\_cm\_global\_validation\_20261029.py} checks the
integrated polynomial budget of \cref{eq:cm-integrand-budget} with
exact rational arithmetic. Independent complex-matrix calculations
check the signed stage-remainder identity, the symmetric defect,
the centered local and global operator remainders, and the sufficient
step-count threshold. Inputs include the Pauli pair, commuting and
zero generators, and generic two- and four-dimensional complex
Hermitian matrices. Direct matrix products and literal finite sums
test the bounds independently of the many-body spectral-kernel code.
The matrix calculations and the 128/256-node quadrature checks test
numerical agreement. They provide neither rigorous enclosures nor
additional many-body step counts.

The accompanying validation runner stages a working copy, checks the
manifest and saved summaries, regenerates the new tables, and reruns
the exact degree-seven certificates and generic-matrix tests. By default it also runs the Casas--Murua global validation driver.
The option \texttt{--comparison-checks} recomputes the five four-site
comparison inputs. Repository-specific paths in
the original scripts and the optional extraction of published
coefficients are documented in the README. The manuscript build
uses the supplied tables and figures and is independent of these
computations.

\subsection{Conventions and verification scope}
The recursive Hall convention in \cref{eq:hall-basis} fixes all signs.
The physical Hamiltonians in \cref{eq:nn-field,eq:power-field} fix the
placement of the random fields. The comparison coefficients come
from~\cite[Appendix~J, Propositions J.1 and J.2]{childs}.
Analytic proofs and exact rational calculations support the stated
identities and inequalities; the many-body norm evaluations and
Casas--Murua quadratures use floating-point arithmetic.
\Cref{sec:formalization,app:lean} give the scope and actual declarations
of the two-term Lean formalization.

\clearpage
\section{Complete three-term degree-five representation}
\label{app:three-coefficients}
For a five-digit word $w=a_1a_2a_3a_4a_5$ define
\begin{equation}
 N_w=[X_{a_1},[X_{a_2},[X_{a_3},[X_{a_4},X_{a_5}]]]].
\end{equation}
The table gives the coefficients of $L_5=\sum_w c_wN_w$ in the verified
right-nested expansion. Omitted coefficients are zero. The words can
be linearly dependent, so this expansion need not be a basis expansion.
We write $144000c_w=a_0+a_1u+a_2u^2$ with integer $a_j$.
The table is generated directly from the rational coefficient data.
\begingroup\small\renewcommand{\arraystretch}{1.13}
\begin{longtable}{crrr@{\hspace{8mm}}crrr}
\caption{Exact three-term quintic coefficients: $144000c_w=a_0+a_1u+a_2u^2$.}
\label{tab:three-full}\\
\toprule $w$ & $a_0$ & $a_1$ & $a_2$ & $w$ & $a_0$ & $a_1$ & $a_2$\\\midrule\endfirsthead
\toprule $w$ & $a_0$ & $a_1$ & $a_2$ & $w$ & $a_0$ & $a_1$ & $a_2$\\\midrule\endhead
\bottomrule\endlastfoot
\texttt{11121} & $-6$ & $52$ & $127$ & \texttt{22232} & $-6$ & $52$ & $127$\\
\texttt{11131} & $-6$ & $52$ & $127$ & \texttt{22321} & $0$ & $0$ & $300$\\
\texttt{11221} & $0$ & $0$ & $300$ & \texttt{22331} & $0$ & $0$ & $300$\\
\texttt{11231} & $0$ & $0$ & $300$ & \texttt{22332} & $0$ & $0$ & $300$\\
\texttt{11232} & $0$ & $0$ & $150$ & \texttt{23121} & $-36$ & $312$ & $-288$\\
\texttt{11321} & $0$ & $0$ & $300$ & \texttt{23131} & $-36$ & $312$ & $-288$\\
\texttt{11331} & $0$ & $0$ & $300$ & \texttt{23221} & $-72$ & $624$ & $-876$\\
\texttt{11332} & $0$ & $0$ & $300$ & \texttt{23231} & $-72$ & $624$ & $-576$\\
\texttt{12121} & $-36$ & $312$ & $-288$ & \texttt{23232} & $-36$ & $312$ & $-288$\\
\texttt{12131} & $-36$ & $312$ & $-288$ & \texttt{23321} & $-72$ & $624$ & $-1176$\\
\texttt{12221} & $-72$ & $624$ & $-976$ & \texttt{23331} & $-72$ & $624$ & $-976$\\
\texttt{12231} & $-72$ & $624$ & $-876$ & \texttt{23332} & $-72$ & $624$ & $-976$\\
\texttt{12232} & $-36$ & $312$ & $-488$ & \texttt{31121} & $12$ & $-104$ & $496$\\
\texttt{12321} & $-72$ & $624$ & $-1176$ & \texttt{31131} & $12$ & $-104$ & $496$\\
\texttt{12331} & $-72$ & $624$ & $-1176$ & \texttt{31221} & $72$ & $-624$ & $1776$\\
\texttt{12332} & $-72$ & $624$ & $-1176$ & \texttt{31231} & $72$ & $-624$ & $1776$\\
\texttt{13121} & $-36$ & $312$ & $-288$ & \texttt{31232} & $36$ & $-312$ & $888$\\
\texttt{13131} & $-36$ & $312$ & $-288$ & \texttt{31321} & $72$ & $-624$ & $1776$\\
\texttt{13221} & $-72$ & $624$ & $-876$ & \texttt{31331} & $72$ & $-624$ & $1776$\\
\texttt{13231} & $-72$ & $624$ & $-576$ & \texttt{31332} & $72$ & $-624$ & $1776$\\
\texttt{13232} & $-36$ & $312$ & $-288$ & \texttt{32121} & $-36$ & $312$ & $-288$\\
\texttt{13321} & $-72$ & $624$ & $-1176$ & \texttt{32131} & $-36$ & $312$ & $-288$\\
\texttt{13331} & $-72$ & $624$ & $-976$ & \texttt{32221} & $0$ & $0$ & $500$\\
\texttt{13332} & $-72$ & $624$ & $-976$ & \texttt{32231} & $0$ & $0$ & $600$\\
\texttt{21121} & $12$ & $-104$ & $496$ & \texttt{32232} & $12$ & $-104$ & $496$\\
\texttt{21131} & $12$ & $-104$ & $496$ & \texttt{32321} & $72$ & $-624$ & $1776$\\
\texttt{21221} & $72$ & $-624$ & $1776$ & \texttt{32331} & $72$ & $-624$ & $1776$\\
\texttt{21231} & $72$ & $-624$ & $1776$ & \texttt{32332} & $72$ & $-624$ & $1776$\\
\texttt{21232} & $36$ & $-312$ & $888$ & \texttt{33121} & $-36$ & $312$ & $-288$\\
\texttt{21321} & $72$ & $-624$ & $1776$ & \texttt{33131} & $-36$ & $312$ & $-288$\\
\texttt{21331} & $72$ & $-624$ & $1776$ & \texttt{33221} & $-72$ & $624$ & $-876$\\
\texttt{21332} & $72$ & $-624$ & $1776$ & \texttt{33231} & $-72$ & $624$ & $-576$\\
\texttt{22121} & $-36$ & $312$ & $-288$ & \texttt{33232} & $-36$ & $312$ & $-288$\\
\texttt{22131} & $-36$ & $312$ & $-288$ & \texttt{33321} & $-24$ & $208$ & $-192$\\
\texttt{22221} & $-24$ & $208$ & $8$ & \texttt{33331} & $-24$ & $208$ & $8$\\
\texttt{22231} & $-24$ & $208$ & $108$ & \texttt{33332} & $-24$ & $208$ & $8$\\
\end{longtable}\endgroup

\clearpage
\section{Degree-five step-count summaries}
\label{app:all-steps}
Each cell is the mean $\pm$ sample standard deviation across five field
realizations. Parameters are $t=n$, $\varepsilon=10^{-3}$. The printed
precision is one decimal place; the CSV retains full numerical precision.
The table values are floating-point evaluations and searches, with the
qualifications in \cref{sec:experiments}.
\begingroup\small\renewcommand{\arraystretch}{1.25}
\begin{longtable}{crrrr}
\caption{Degree-five mean step counts and sample standard deviations.}\\
\toprule $n$ & Paper & Local + remainder & Cancellation & Empirical\\\midrule\endfirsthead
\toprule $n$ & Paper & Local + remainder & Cancellation & Empirical\\\midrule\endhead
\midrule\multicolumn{5}{r}{Continued on next page}\\\endfoot
\bottomrule\endlastfoot
\multicolumn{5}{l}{\textbf{Nearest-neighbor, even--odd}}\\\midrule
4 & $116.0 \pm 4.8$ & $48.6 \pm 2.2$ & $48.2 \pm 2.2$ & $24.0 \pm 0.7$\\
5 & $197.2 \pm 4.0$ & $83.6 \pm 1.5$ & $82.0 \pm 1.7$ & $41.4 \pm 0.5$\\
6 & $262.8 \pm 6.3$ & $111.4 \pm 2.9$ & $108.8 \pm 2.7$ & $52.6 \pm 0.5$\\
7 & $338.8 \pm 7.7$ & $142.2 \pm 3.6$ & $138.6 \pm 3.4$ & $71.2 \pm 2.3$\\
8 & $409.2 \pm 8.2$ & $168.6 \pm 4.8$ & $163.8 \pm 4.2$ & $86.0 \pm 2.9$\\
9 & $503.2 \pm 2.4$ & $202.2 \pm 2.2$ & $195.2 \pm 1.6$ & $109.0 \pm 3.3$\\
10 & $585.4 \pm 7.2$ & $231.8 \pm 2.9$ & $223.0 \pm 2.1$ & $123.8 \pm 0.8$\\
\multicolumn{5}{l}{\textbf{Power law, XYZ, $\alpha=0$}}\\\midrule
4 & $380.6 \pm 9.0$ & $124.0 \pm 2.4$ & $124.0 \pm 2.4$ & $46.8 \pm 0.4$\\
5 & $738.6 \pm 21.6$ & $232.6 \pm 5.7$ & $232.6 \pm 5.7$ & $84.2 \pm 0.4$\\
6 & $1351.4 \pm 14.8$ & $405.6 \pm 4.2$ & $405.2 \pm 4.1$ & $143.0 \pm 1.9$\\
7 & $2048.8 \pm 8.0$ & $614.2 \pm 2.5$ & $614.2 \pm 2.5$ & $217.4 \pm 0.9$\\
8 & $3064.0 \pm 12.6$ & $928.2 \pm 2.3$ & $928.2 \pm 2.3$ & $317.8 \pm 3.6$\\
9 & $4181.0 \pm 33.7$ & $1247.6 \pm 10.0$ & $1247.6 \pm 10.0$ & $442.6 \pm 4.0$\\
10 & $5623.0 \pm 21.3$ & $1682.6 \pm 4.1$ & $1682.6 \pm 4.1$ & $596.2 \pm 6.0$\\
\multicolumn{5}{l}{\textbf{Power law, XYZ, $\alpha=4$}}\\\midrule
4 & $192.0 \pm 13.6$ & $68.0 \pm 3.7$ & $67.0 \pm 3.7$ & $28.0 \pm 0.0$\\
5 & $295.4 \pm 14.3$ & $103.6 \pm 3.8$ & $102.2 \pm 4.2$ & $42.4 \pm 1.8$\\
6 & $398.0 \pm 21.9$ & $137.6 \pm 6.0$ & $135.4 \pm 6.2$ & $56.4 \pm 2.2$\\
7 & $496.6 \pm 14.8$ & $169.8 \pm 4.3$ & $166.8 \pm 4.1$ & $72.6 \pm 4.0$\\
8 & $618.4 \pm 25.3$ & $207.0 \pm 7.8$ & $202.4 \pm 7.8$ & $91.4 \pm 4.4$\\
9 & $745.8 \pm 20.2$ & $244.8 \pm 5.2$ & $238.6 \pm 5.5$ & $113.4 \pm 4.9$\\
10 & $873.4 \pm 22.7$ & $282.4 \pm 5.3$ & $274.2 \pm 5.2$ & $129.0 \pm 2.4$\\
\end{longtable}\endgroup

\clearpage
\section{Refined finite-size step-count summaries}
\label{app:refined-steps}
We evaluate the refined bound in \cref{eq:septic-norm} on all 105
instances.
Each entry is the mean $\pm$ sample standard deviation over five fields.
The degree-five column is the cancellation bound from \cref{sec:global};
the refined column includes the grouped ninth-order remainder and all
nonlinear product terms. The empirical column is unchanged.
\begingroup\small\renewcommand{\arraystretch}{1.25}
\begin{longtable}{crrrr}
\caption{Refined mean step counts and sample standard deviations.}\\
\toprule $n$ & Paper & Degree-five & Refined & Empirical\\\midrule\endfirsthead
\toprule $n$ & Paper & Degree-five & Refined & Empirical\\\midrule\endhead
\midrule\multicolumn{5}{r}{Continued on next page}\\\endfoot
\bottomrule\endlastfoot
\multicolumn{5}{l}{\textbf{Nearest-neighbor, even--odd}}\\\midrule
4 & $116.0 \pm 4.8$ & $48.2 \pm 2.2$ & $26.2 \pm 0.8$ & $24.0 \pm 0.7$\\
5 & $197.2 \pm 4.0$ & $82.0 \pm 1.7$ & $45.2 \pm 0.8$ & $41.4 \pm 0.5$\\
6 & $262.8 \pm 6.3$ & $108.8 \pm 2.7$ & $58.8 \pm 1.3$ & $52.6 \pm 0.5$\\
7 & $338.8 \pm 7.7$ & $138.6 \pm 3.4$ & $77.6 \pm 3.0$ & $71.2 \pm 2.3$\\
8 & $409.2 \pm 8.2$ & $163.8 \pm 4.2$ & $93.2 \pm 4.4$ & $86.0 \pm 2.9$\\
9 & $503.2 \pm 2.4$ & $195.2 \pm 1.6$ & $114.4 \pm 3.6$ & $109.0 \pm 3.3$\\
10 & $585.4 \pm 7.2$ & $223.0 \pm 2.1$ & $129.4 \pm 1.1$ & $123.8 \pm 0.8$\\
\multicolumn{5}{l}{\textbf{Power law, XYZ, $\alpha=0$}}\\\midrule
4 & $380.6 \pm 9.0$ & $124.0 \pm 2.4$ & $47.0 \pm 0.0$ & $46.8 \pm 0.4$\\
5 & $738.6 \pm 21.6$ & $232.6 \pm 5.7$ & $84.8 \pm 0.8$ & $84.2 \pm 0.4$\\
6 & $1351.4 \pm 14.8$ & $405.2 \pm 4.1$ & $143.2 \pm 1.9$ & $143.0 \pm 1.9$\\
7 & $2048.8 \pm 8.0$ & $614.2 \pm 2.5$ & $218.2 \pm 0.4$ & $217.4 \pm 0.9$\\
8 & $3064.0 \pm 12.6$ & $928.2 \pm 2.3$ & $318.4 \pm 3.0$ & $317.8 \pm 3.6$\\
9 & $4181.0 \pm 33.7$ & $1247.6 \pm 10.0$ & $443.2 \pm 3.6$ & $442.6 \pm 4.0$\\
10 & $5623.0 \pm 21.3$ & $1682.6 \pm 4.1$ & $597.2 \pm 6.0$ & $596.2 \pm 6.0$\\
\multicolumn{5}{l}{\textbf{Power law, XYZ, $\alpha=4$}}\\\midrule
4 & $192.0 \pm 13.6$ & $67.0 \pm 3.7$ & $32.8 \pm 0.8$ & $28.0 \pm 0.0$\\
5 & $295.4 \pm 14.3$ & $102.2 \pm 4.2$ & $50.0 \pm 1.4$ & $42.4 \pm 1.8$\\
6 & $398.0 \pm 21.9$ & $135.4 \pm 6.2$ & $65.6 \pm 1.9$ & $56.4 \pm 2.2$\\
7 & $496.6 \pm 14.8$ & $166.8 \pm 4.1$ & $82.0 \pm 3.2$ & $72.6 \pm 4.0$\\
8 & $618.4 \pm 25.3$ & $202.4 \pm 7.8$ & $100.8 \pm 4.2$ & $91.4 \pm 4.4$\\
9 & $745.8 \pm 20.2$ & $238.6 \pm 5.5$ & $121.6 \pm 4.4$ & $113.4 \pm 4.9$\\
10 & $873.4 \pm 22.7$ & $274.2 \pm 5.2$ & $138.0 \pm 2.3$ & $129.0 \pm 2.4$\\
\end{longtable}\endgroup

\end{document}